\documentstyle[12pt]{article}
\begin{document}

\newtheorem{theoreme}{{\bf Th\'eor\`eme}}[subsection]
\newtheorem{proposition}[theoreme]{{\bf Proposition}}
\newtheorem{lemme}[theoreme]{{\bf Lemme}}
\newtheorem{corollaire}[theoreme]{{\bf Corollaire}}
\newtheorem{definition}[theoreme]{{\bf D\'efinition}}

\def\sousection{\noindent
\addtocounter{subsection}{1}{\bf\arabic{section}.\arabic{subsection}~~}}

\font\huit=cmr8
\font\neuf=cmr9
\font\dix=cmr10
\font\goth=eufm10
\font\huit=cmr8
\font\neuf=cmr9
\font\bb=msym10

\def\dsp{\displaystyle}

\def\calm{\cal M}
\def\get{\hbox{{\goth g}}}
\def\sal{\hbox{\goth s}}
\def\g{\gamma}
\def\om{\omega}
\def\r{\rho}
\def\a{\alpha}
\def\s{\sigma}
\def\vfi{\varphi}
\def\l{\lambda}
\def\implique{\Rightarrow}
\def\o{{\circ}}
\def\Diff{\hbox{\rm Diff}}
\def\S1{\hbox{\rm S$^1$}}
\def\Vect{\hbox{\rm Vect}}
\def\ad{\hbox{\hbox{\rm \Ad}}}
\def\semid{\hbox{\bb o}}
\def\blanc{\hbox{\ \ }}
\def\To{\hbox{\bb T}}
\def\Gt{\widetilde{G}}
\def\gtilde{\tilde{g}}
\def\Ad{\hbox{\rm Ad}}
\def\ad{\hbox{\rm ad}}
\def\interieur#1{\displaystyle{\buildrel\,\,\circ\over#1}}
\def\semid{\hbox{\bb o}}

\def\abstracta{\if@twocolumn
\section*{Abstract}
\else \small
\begin{center}
{\bf Abstract\vspace{-.5em}\vspace{0pt}}
\end{center}
\quotation
\fi}
\def\endabstracta{\if@twocolumn\else\endquotation\fi}

\def\pds#1,#2{\langle #1\mid #2\rangle} 
\def\f#1,#2,#3{#1\colon#2\to#3} 

\def\hfl#1{{\buildrel{#1}\over{\hbox to
12mm{\rightarrowfill}}}}
\def\ghfl#1{{\buildrel{#1}\over{\hbox to
12mm{\leftarrowfill}}}}
\def\vfl#1{\llap{$\scriptstyle#1$}\left\downarrow\vbox to
6mm{}\right.}
\def\vflup#1{\llap{$\scriptstyle#1$}\left\uparrow\vbox to
6mm{}\right.} 
\def\diagram#1{\def\normalbaselines{\baselineskip=0pt
\lineskip=5pt\lineskiplimit=1pt} \matrix{#1}}

\font\twelve=cmbx10 at 15pt
\font\ten=cmbx10 at 12pt
\font\eight=cmr8
\renewcommand{\thefootnote}{\fnsymbol{footnote}}

\begin{titlepage}

\begin{center}

{\ten Centre de Physique Th\'eorique\footnote{Unit\'e Propre de
Recherche 7061} - CNRS - Luminy, Case 907}

{\ten F-13288 Marseille Cedex 9 - France }

\vspace{1 cm}
\setcounter{footnote}{0}
\renewcommand{\thefootnote}{\arabic{footnote}}
\twelve STRUCTURES SYMPLECTIQUES SUR LES ESPACES

\twelve DE COURBES PROJECTIVES ET AFFINES

\vspace{0.3 cm}

{\bf L. GUIEU}\footnote{e-mail~: guieu@cptvax.in2p3.fr} {\bf et V.Yu.
OVSIENKO}\footnote{e-mail~: ovsienko@cptvax.in2p3.fr}

\vspace{1 cm}

{\bf R\'esum\'e}
\end{center}
Nous mettons en \'evidence l'existence d'une structure symplectique
sur l'espace des courbes non param\'etr\'ees et non
d\'eg\'en\'er\'ees d'une vari\'et\'e localement affine. Nous
consid\'erons \'egalement des espaces de courbes projectives munis
d'une structure de Poisson. Cette cons\-truction relie l'alg\`ebre de
Virasoro et le crochet de Gel'fand-Dikii \`a la g\'eom\'etrie
projective diff\'erentielle.

\bigskip

\centerline{\bf Abstract}

\medskip

A symplectic structure on the space of nondegenerate and
nonparametrized curves in a locally affine manifold is defined. We
also consider several interesting spaces of nondegenerate projective
curves endowed with Poisson structures. This construction connects
the Virasoro algebra and the Gel'fand-Dikii bracket with the
projective differential geometry.

\vskip 1truecm

\noindent Classification AMS (1991) : 54B20 58F05 17B68

\bigskip

\noindent Mots-cl\'es : G\'eom\'etrie diff\'erentielle projective -
Espaces de courbes - Alg\`ebre de Virasoro - Crochet de Gel'fand-Dikii
\medskip
\noindent Janvier 1994

\noindent CPT-94/P.2980

\bigskip

\noindent anonymous ftp or gopher: cpt.univ-mrs.fr

\end{titlepage}

\section{INTRODUCTION}
Il existe beaucoup de fa\c cons de construire des structures symplectiques sur
 les espaces de courbes dans des vari\'et\'es.

La vari\'et\'e dont on consid\`ere l'espace de courbes doit \^etre munie d'une
certaine structure g\'eom\'etrique. De nombreux travaux ont \'et\'e d\'edi\'es
au cas
des vari\'et\'es riemanniennes (voir par exemple [Ig][Iv]) et symplectiques.
Enfin, le cas des vari\'et\'es de dimension 3 munies
d'une forme volume a \'et\'e consid\'er\'e dans [Br].

Nous nous int\'eresserons, dans cet article, \`a l'espace des courbes non
d\'eg\'en\'er\'ees\footnote{c'est-\`a-dire qu'en chacun de ses points, son
drapeau
osculateur est non d\'eg\'en\'er\'e.} sur une vari\'et\'e localement
projective. La
question principale qui est \`a l'origine de ce travail est la suivante~:
\vskip0.3cm
{\tenrm Existe-t-il une forme symplectique sur cet espace qui ne d\'epende que
de
la structure projective plac\'ee sur la vari\'et\'e~?}
\vskip0.3cm
Cette question nous am\`ene naturellement \`a la th\'eorie des alg\`ebres de
Lie de
dimension infinie. Pr\'ecisons ce fait par un premier exemple~: la forme
symplectique
sur l'espace des structures projectives du cercle unit\'e $S^1$ est reli\'ee
\`a la
structure de Poisson canonique dont est dot\'e le dual de l'alg\`ebre de
Virasoro--. C'est
la g\'en\'eralisation de cette structure -- le crochet de Gel' fand - Dikii --
qui
prend le relais dans le cas d'une vari\'et\'e projective de dimension $>1$. Le
lien
existant entre la g\'eom\'etrie diff\'erentielle projective et les alg\`ebres
de Lie
de dimension infinie constitue le sujet principal de notre article.

Les r\'esultats nouveaux sont les suivants~:
\vskip0.3cm
\sousection  Nous pr\'esentons une d\'efinition purement g\'eom\'etrique de la
forme
symplectique sur l'espace de toutes les structures projectives sur $S^1$ \`a
monodromie fix\'ee (chap. 2.7). Cette 2-forme co\"{\i}ncide avec la forme de
Kirillov
- Kostant - Souriau sur les orbites coadjointes dans le dual de l'alg\`ebre de
Virasoro $Vir$ (voir [Ki]).
\vskip0.3cm
\sousection  Soit $\Gamma_T$ l'espace des courbes $\gamma$ param\'etr\'ees et
non
d\'eg\'en\'er\'ees sur la sph\`ere $S^n$ \`a monodromie $T\in
PGL(n+1\;;\;I\!\!R)$ fix\'ee (c'est-\`a-dire~:
$\gamma(x + 2\pi)= T. \gamma(x)\;\;\;\forall x\in I\!\!R$). Notons
$$\displaystyle\overline{\Gamma}_T = \frac{\Gamma_T}{PGL(n+1\;;\;I\!\!R)}$$ le
quotient
de cet espace par la relation d'\'equivalence projective. Nous d\'emontrons
alors le~:
\vskip0.3cm
{\bf Th\'eor\`eme.~}{\it (i) [KO,O] Les espaces $\overline{\Gamma}_T$ sont les
feuilles symplectiques du crochet de Adler-Gel'fand-Diki\"{\i}.}

{\it (ii) L'application du moment
$$\mu :\overline{\Gamma}_T\rightarrow (Vir)^{\star}$$
pour l'action naturelle du groupe $\Diff(S^1)$ des
diff\'eomorphismes du cercle, co\"{\i}ncide avec la courbure projective.}
\vskip0.3cm
{\bf Corollaire:~}{\it Il existe sur $\overline{\Gamma}_T$ une forme
symplectique invariante sous l'action du groupe  $\Diff(S^1)$ .} \vskip0.3cm

Nous
pr\'esenterons dans le chapitre 7.4 la formule pr\'ecise de la forme
symplectique construite sur l'espace des courbes non oscillantes de
$I\!\!RP^2$.

Toutefois, nous nous attacherons plut\^ot \`a examiner le cas de l'espace des
courbes
{\it g\'eom\'etriques} (c'est-\`a-dire~: non param\'etr\'ees), la structure
symplectique induite par le crochet de Adler-Gel' fand - Dikii servant de fil
d'Ariane. Cette
forme symplectique sera calcul\'ee explicitement dans le cas, plus simple, des
courbes dans le plan projectif.
\vskip0.3cm
\sousection
Soit $M$ une vari\'et\'e localement affine de dimension $n$. La notation ${\cal
C}_{\mbox{id}}(M)$ d\'esignera l'espace des courbes ferm\'ees, non
d\'eg\'en\'er\'ees
et {\it non param\'etr\'ees} sur $M$ et $\overline{{\cal C}}_{\mbox{id}}(M)$
l'espace
quotient de ${\cal C}_{\mbox{id}}(M)$ par la relation d'\'equivalence affine
des
courbes. Nous avons alors le~:
\vskip0.3cm
{\bf Th\'eor\`eme 1~}{\it
$\overline{{\cal C}}_{id}(M)$ est une vari\'et\'e symplectique dont chaque
composante
connexe est isomorphe \`a un certain espace $\overline{\Gamma}_T(S^{n-1})$.
}
\vskip0.3cm

Remarquons que cet isomorphisme permet de d\'efinir une action du groupe
$\Diff(\S1)$
sur l'espace des courbes non param\'etr\'ees~!
\vskip0.3cm
\sousection
En ce qui concerne le cas plus int\'eressant de l'espace des courbes sur une
vari\'et\'e localement projective, une construction analogue au cas affine ne
conduit
pas \`a l'existence d'une structure symplectique mais \`a celle d'un crochet de
Poisson. De plus, la d\'efinition de la non d\'eg\'en\'erescence doit \^etre
renforc\'ee: dans le cas de $I\!\!R P^2$, les courbes doivent \^etre non
d\'eg\'en\'er\'ees (au
sens classique) mais doivent \^etre \'egalement exemptes de points
sextactiques\footnote{Cette terminologie peut s'expliquer par cette remarque de
Elie
Cartan: ``La c\^onique osculatrice passant par un point sextactique a six
points au
moins en commun avec la courbe.''}: ce sont ceux en lesquels la courbe
entretient
avec sa c\^onique osculatrice un contact d'ordre $\geq 5$.

Ces courbes seront appel\'ees {\it courbes de Cartan}.
\vskip0.3cm
{\bf Th\'eor\`eme 2~}{\it
L'espace des courbes de Cartan non param\'etr\'ees et \`a monodromie fix\'ee
est une
vari\'et\'e de Poisson~: il existe une immersion de cet espace dans l'espace
des
structures projectives sur $S^1$
}
\vskip0.3cm
\sousection
En g\'en\'eral, sur l'espace des courbes de $I\!\!R P^2$ non param\'etr\'ees,
non
d\'eg\'en\'er\'ees et \`a monodromie fix\'ee (sans conditions sur
l'\'eventuelle
pr\'esence de points sextactiques), nous d\'efinissons un {\it feuilletage
symplectique}~: chaque feuille est munie d'une d'une forme symplectique. Les
courbes
qui sont sur une m\^eme feuille poss\`edent le m\^eme nombre de points
sextactiques.

Un probl\`eme d'ordre topologique qui nous semble int\'eressant se pose alors~:
\vskip0.3cm
{\bf Probl\`eme:~}{\it Classer \`a homotopie pr\`es les courbes dans $I\!\!R
P^2$ non d\'eg\'en\'er\'ees, \`a monodromie fix\'ee et poss\'edant $k$ points
sextactiques.}
 \vskip0.3cm
Par exemple, pour les courbes ferm\'ees~: $k$ est pair $\geq 6$ (il s'agit d'un
th\'eor\`eme classique de g\'eom\'etrie diff\'erentielle affine~: le
th\'eor\`eme des
6 sommets. Voir [Bo][Bu] et le chapitre 2). Notons que la classification des
courbes ferm\'ees et non d\'eg\'en\'er\'ees (sans conditions sur le nombre de
points sextactiques) a \'et\'e effectu\'ee dans [Li1] (voir aussi chap. 2).

Toutes les constructions de cet article font intervenir deux structures
g\'eom\'etriques sur
une courbe~: sa longueur (affine ou projective) et sa courbure (affine ou
projective).

\vfill\eject

\section{El\'ements de la g\'eom\'etrie diff\'erentielle des courbes
projectives et
affines}

Nous rassemblons ici quelques r\'esultats de la g\'eom\'etrie diff\'erentielle
r\'eelle des courbes dans l'espace projectif (ou affine).

La g\'eom\'etrie projective \'etudie les propri\'et\'es projectivement
invariantes de
ces objets. Le groupe $PGL (n\;;\;I\!\!R)$ des ``sym\'etries projectives'' ne
laisse
sur l'espace $I\!\!RP^{n-1}$ aucune structure invariante (riemannienne,
symplectique, de contact $\ldots$ etc) mis \`a part la structure projective.
C'est la
raison pour laquelle, les notions les plus \'el\'ementaires de la g\'eom\'etrie
diff\'erentielle projective ne sont gu\`ere ais\'ees \`a d\'ecrire.

Nous pr\'esentons dans ce chapitre des r\'esultats classiques (qui, sans
doutes,
m\'eriteraient d'\^etre plus connus) sous une forme contemporaine et de
mani\`ere
\'el\'ementaire.

\subsection{Structures projectives et affines sur une vari\'et\'e}

Une structure projective sur une vari\'et\'e est une fa\c con d'identifier
(localement et en chacun de ses points) cette vari\'et\'e avec l'espace
projectif.

\begin{definition}
\hfill\break
(i) Soit $M$ une vari\'et\'e diff\'erentiable de dimension $n$. Un atlas
projectif
sur $M$ est la donn\'ee d'une famille de plongements $\varphi _\alpha \;:
U_\alpha \rightarrow I\!\!RP ^n\;\;\;(\alpha \in I)$ dont les domaines de
d\'efinition forment un recouvrement ouvert de $M$ et telle que les changements
de
cartes $g_{\alpha \beta } = \varphi _\alpha \circ \varphi _\beta ^{-1}$ soient
des
restrictions d'\'el\'ements du groupe projectif $PGL(n+1\;;\;I\!\!R)$.
\vskip0.4cm
(ii) Deux atlas projectifs sont dits {\rm \'equivalents} si leur r\'eunion est
encore
un atlas projectif.
\vskip0.4cm
(iii) Une {\rm structure projective} sur $M$ est la donn\'ee d'une classe
d'\'equivalence d'atlas projectifs.
\end{definition}

{\bf Exemple:~}
Toute surface orientable admet des structures projectives.
\vskip0.3cm
{\bf D\'efinition -- Groupe de monodromie~:}
Une structure projective $\sigma $ sur $M$ d\'efinit (modulo conjugaison) une
immersion du rev\^etement universel de $M$ dans l'espace projectif de dimension
$n$~:
$$\varphi \;:\; \widetilde M \rightarrow I\!\!R P^n $$
ainsi qu'un homomorphisme du groupe fondamental de $M$ dans le groupe
projectif~:
$$ T\;: \pi_1 (M) \rightarrow PGL(n+1\;;\;I\!\!R).$$
$\varphi $ et $T$ devant satisfaire la relation d'\'equivariance suivante~:
$$\varphi (a.\xi ) = T(a). \varphi (\xi )\hskip1cm\forall a \in \pi_1(M),
\forall \xi
\in \widetilde M.$$
$\varphi $ est appel\'ee une {\it application d\'eveloppante} pour la structure
$\sigma $ et
l'image de $T$ est le {\it groupe de monodromie} associ\'e \`a $\varphi $.
\vskip0.3cm
Le couple $(\varphi, T)$ est d\'efini modulo l'action du groupe projectif~:

$$A. (\varphi ,T) = (A. \varphi , A. T. A^{-1})\;\;\;\;A\in PGL(n+1\;;
I\!\!R)$$
$(\varphi ,T)$ est une paire d\'eveloppante pour la structure $\sigma $.

\begin{definition}
Une structure projective sur une vari\'et\'e $M$ est appel\'ee une {\rm
structure
affine} si tous les changements de cartes $g_{\alpha \beta }$ sont dans le
groupe
affine~:
$$Aff(n\;;\;I\!\!R) \cong GL(n\;;\;r)\semid I\!\!R^n \hookrightarrow PGL(n+1\;;
I\!\!R)$$
(\`a conjugaison pr\`es).
\end{definition}

Les applications d\'eveloppantes d'une structure affine sont \`a valeurs dans
$I\!\!R^n$:
$$\varphi \;:\; \widetilde M \rightarrow I\!\!R^n$$

\subsection{Espace des courbes non d\'eg\'en\'er\'ees}

$M$ d\'esigne ici une vari\'et\'e localement projective de dimension $n$,
c'est-\`a-dire un couple form\'e par une vari\'et\'e et une structure
projective sur
cette derni\`ere.

\begin{definition}Une courbe ferm\'ee $\gamma\;: S^1 \rightarrow M$ est dite
{\rm non d\'eg\'en\'er\'ee} si, dans une carte projective, les vecteurs
$\gamma ',\gamma'',\ldots,\gamma^{(n)}$ sont lin\'eairement ind\'ependants.
\end{definition}

{\bf Remarque~}
Le {\it drapeau} engendr\'e par ces vecteurs est non d\'eg\'en\'er\'e et
ind\'ependant du choix d'une carte projective.
\vskip0.3cm
{\bf Notion d'\'equivalence projective - D\'efinition~:}

(i) Deux courbes $\gamma_1$ et $\gamma_2$ dans $I\!\!R P^n$ seront dites
{\it projectivement equivalentes} si il existe $A\in PGL(n+1\;; I\!\!R)$ tel
que~:
$$\gamma_2 = A\circ \gamma_1$$

(ii) Soit $M$ une vari\'et\'e localement projective de paire d\'eveloppante
$(\varphi
,T)$.
Une {\it d\'evelopp\'ee} d'une courbe $\gamma\;: I\!\!R\rightarrow M$ est une
courbe
$I\!\!R\rightarrow I\!\!R P^n$ de la forme $\varphi \circ \widetilde\gamma$,
o\`u
$\widetilde\gamma$ d\'esigne un rel\`evement de $\gamma$ dans le rev\^etement
universel de $M$.
\vskip0.5cm
(iii) Deux courbes $\gamma_1$ et $\gamma_2$ dans $M$ seront dites {\it
projectivement
\'equivalentes} si elles poss\`edent deux d\'evelopp\'ees qui le sont. Cette
d\'efinition est ind\'ependante du choix de ces d\'evelopp\'ees.

\vskip0.5cm
Notons que cette d\'efinition s'\'etend sans probl\`emes au cas des courbes non
param\'etr\'ees.
\vskip0.3cm
{\bf Equivalence affine~}
La notion d'\'equivalence des courbes dans une vari\'et\'e localement affine se
d\'efinit de mani\`ere analogue~: il suffit de remplacer $I\!\!R P^n$ par
$I\!\!R^n$
et $PGL(n+1\;; I\!\!R)$ par $Aff(n\;; I\!\!R)$.
\vskip0.3cm
Nous garderons la m\^eme notation pour les espaces de courbes modulo
\'equivalence
affine.
\vskip0.3cm
{\bf Cas des courbes \`a monodromie~}
Soit $\gamma$ une courbe projective non d\'eg\'en\'er\'ee dans $I\!\!R P^n$ et
$T$ un
\'el\'ement de $PGL(n+1\;; I\!\!R)$. Nous dirons que cette courbe est {\it
T-\'equivariante} si~:
$$\gamma(x + 2\pi) = T. \gamma(x)\;\;\;\; \forall x \in I\!\!R$$

G\'eom\'etriquement, il est plus important d'\'etudier les courbes \`a
\'equivalence
projective pr\`es. La monodromie est alors bien d\'efinie si on la consid\`ere
comme
une classe de conjugaison dans $PGL(n+1\;; I\!\!R)$.
Pour le cas affine, les d\'efinitions sont analogues.
\vskip0.3cm
{\bf Notations~}
On notera~:

(i) ${\cal P}(M)$ l'espace de toutes les structures projectives sur une
vari\'et\'e
$M$
\vskip0.4cm
(ii) ${\cal P}_T(M)\subset {\cal P}(M)$ l'espace des structures projectives sur
$M$
dont la monodromie est fix\'ee et \'egale \`a $T$ ($T$ d\'esigne ici un
homomorphisme
$T\;: \pi_1(M) \rightarrow PGL(n+1\;; I\!\!R)$ d\'efini \`a conjugaison
pr\`es).
\vskip0.3cm
(iii) ${\cal C}(M)$ l'espace de toutes les courbes non d\'eg\'en\'er\'ees et
non
param\'etr\'ees sur une vari\'et\'e $M$ munie d'une structure projective (ou
affine).
\vskip0.3cm
(iv) ${\cal C}_0(M)\subset {\cal C}(M)$ l'espace des courbes \'equivariantes
(la
monodromie n'est pas fix\'ee).
\vskip0.3cm
(v) ${\cal C}_T(M)\subset {\cal C}_0(M)$ l'espace des courbes \'equivariantes
\`a
monodromie fix\'ee $= T$

($T$ d\'esignant ici une classe de conjugaison dans $PGL(n+1\;; I\!\!R)$).
\vskip0.3cm
(vi) $\overline{{\cal C}}(M), \overline{{\cal C}}_0(M)\;\mbox{ et }\;
\overline{{\cal
C}}_T(M)$ d\'esignent les espaces quotients par la relation d'\'equivalence
projective (ou affine) des espaces d\'efinis ci-dessus.
\vskip0.3cm
(vii) $\Gamma(M)$ l'espace des courbes non d\'eg\'en\'er\'ees et {\it
param\'etr\'ees}.
\vskip0.3cm
Les espace $\Gamma_0(M), \Gamma_T(M), \overline{\Gamma}(M)$ etc$\ldots$ se
d\'efinissent de la m\^eme fa\c con.

\subsection{Homotopie.}

Une question d'ordre topologique se pose imm\'ediatement, \`a savoir~: calculer
les
classes d'homotopie de courbes non-d\'eg\'en\'er\'ees sur une vari\'et\'e
localement
projective (ou localement affine). Il s'agit ici d'homotopie \`a travers les
courbes
{\it ferm\'ees} et {\it non d\'eg\'en\'er\'ees}. La question est donc de
d\'eterminer
les composantes connexes de l'espace ${\cal C}_{id}(M)$. Nous pr\'esentons ici
des
r\'esultats connus~:
\vskip0.3cm
{\bf Composantes connexes de ${\cal C}_{id}(I\!\!R^2)$~}
Ces classes d'homotopie sont rep\'er\'ees par un entier~: le degr\'e de
l'application
de Gauss (voir [Li1] [Wh]) associ\'ee \`a chaque courbe. Il y en a donc un
nombre
infini d\'enombrable.
\vskip0.3cm
{\bf Le th\'eor\`eme de Little [L1]~}
{\it L'espace ${\cal C}_{id}(I\!\!R P^2)$ poss\`ede trois composantes
connexes~: toute
courbe ferm\'ee et non d\'eg\'en\'er\'ee sur $I\!\!R P^2$ est homotope \`a
l'une des
trois courbes dessin\'ees ci-dessous (fig. 1)~:}
\vskip6cm
\centerline {\bf Fig. 1}
\vskip0.3cm
Remarquons que, dans la classe des courbes lisses, les courbes (a) et (c) sont
isomorphes.
\vskip0.3cm
L'espace ${\cal C}_{id}(I\!\!R^4)$ poss\`ede encore trois composantes connexes
chacune correspondant \`a l'une des courbes de la Fig. 1 via l'application de
Gauss
(voir [L2]).

\vskip0.3cm
La classification des composantes connexes de l'espace ${\cal C}_{T}(I\!\!R
P^n)$
pour tout $T\in PGL(3\;; I\!\!R)$ est donn\'ee dans [K-S].

\vskip0.3cm
Les classes d'homotopie des courbes sur $I\!\!R P^n$ ont \'et\'e calcul\'ees
r\'ecemment dans [Sh] o\`u il est d\'emontr\'e que l'espace ${\cal
C}_{id}(I\!\!R
P^n)$ poss\`ede 3 composantes connexes pour $n$ pair et 2 pour $n$ impair.

\subsection{Courbes projectives et op\'erateurs diff\'erentiels lin\'eaires}

Notons ${\cal L}^n$ l'espace des op\'erateurs diff\'erentiels lin\'eaires $L$
d'ordre
$n \geq 2$ de la forme~:
$$L = \partial^n + u_{n-2}(x) \partial^{n-2} + u_{n-3}(x) \partial^{n-3}
+\ldots +
u_0(x)\eqno {(1)}$$
o\`u $\displaystyle\partial = \frac{d}{dx}$ et $u_i \in C^\infty
(I\!\!R)\;\;\;\;
\forall i \in \{ 0, 1, \ldots, n-2\}$.

Le fait suivant pr\'ecise le rapport existant entre les op\'erateurs $(1)$ et
les
courbes (param\'etr\'ees) non d\'eg\'en\'er\'ees sur $I\!\!R P^{n-1}$~:

\begin{lemme}[{Wilczynski [W], [C]}]
L'espace des courbes projectives (param\'etr\'ees) non d\'eg\'en\'er\'ees
modulo
\'equivalence projective est isomorphe \`a l'espace des op\'erateurs
diff\'erentiels
lin\'eaires (1):
$$\overline\Gamma (I\!\!R P^{n-1})\cong {\cal L}^n\eqno{(2)}$$
\end{lemme}

\noindent{\bf Preuve~:}
\vskip0.3cm
(i) On associe \`a chaque op\'erateur de la forme (1) une classe de courbes
projectivement \'equivalentes dans $I\!\!R P^{n-1}$ par la construction
suivante~:

Soit $E$ l'espace vectoriel de dimension $n$ des solutions de l'\'equation
diff\'erentielle $L y = 0$. A chaque $x\in I\!\!R$ correspond l'hyperplan $H_x
\subset E$ constitu\'e des solutions s'annulant en $x$~:

$$ H_x = \{ y\in E \vert y(x) = 0 \}$$

L'application $x\in I\!\!R \mapsto H_x \in P(E^\ast)\cong I\!\!R P^{n-1}$
fournit la
courbe $\gamma(x)$ recherch\'ee.

Les deux faits suivants sont alors \'evidents~: la courbe $\gamma$ est
d\'efinie \`a
homographie pr\`es (correspondant au choix d'une base de $E$) et elle est non
d\'eg\'en\'er\'ee.
\vskip0.3cm
(ii) R\'eciproquement, soit $\gamma\in \Gamma(I\!\!R P^{n-1})$ il existe un
unique
op\'erateur $L_\gamma \in {\cal L}^n$ dont la classe de courbes associ\'ee par
la
construction pr\'ec\'edente soit la classe d'\'equivalence $\overline\gamma
\in\overline\Gamma(I\!\!R P^{n-1})$. La d\'efinition exacte de l'op\'erateur
$L_\gamma$ sera d\'ecrite dans la d\'emonstration du lemme 2.5.3
ci-apr\`es.$\Box$
\vskip0.3cm
{\bf Remarques~:}

(a) {\it Monodromie~:} Soit $\gamma\in \Gamma_T(I\!\!R P^{n-1})$ une courbe
T-\'equivariante (o\`u $T$ est une classe de conjugaison dans $PGL(n\;;
I\!\!R)$).
Les coefficients $u_i$ dans l'op\'erateur $L_\gamma$ sont alors
$2\pi$-p\'eriodiques. L'op\'erateur de monodromie $\widehat T$ associ\'e \`a
$L_\gamma$~:
$$\widehat T\;: E \rightarrow E\;: \; y\mapsto y\circ \tau _{2\pi}\;\;\;\;(\tau
_{2\pi}(x) = x + 2\pi)$$
est alors un \'el\'ement de $SL(E)$. La donn\'ee de $\widehat T$ est
\'equivalente
\`a celle d'une classe de conjugaison dans $SL(n\;; I\!\!R)$~; celle-ci se
projette
sur $T$~:

$$SL(n\;; I\!\!R)\rightarrow PGL(n\;; I\!\!R)\;\;\;\; \widehat T\mapsto T$$

(b) Tous les {\it invariants diff\'erentiels projectifs} des courbes $\gamma
\in
\Gamma (I\!\!R P^{n-1})$ s'expriment comme des polyn\^omes diff\'erentiels en
les
fonctions $u_i$.
\vskip0.3cm
{\bf Exemples~}
Un op\'erateur de {\it Sturm-Liouville~}:
$$L = \partial^2 + u(x)$$
dont le  potentiel $u$ est $2\pi$-p\'eriodique est appel\'e un {\it
op\'erateur de Hill}.

\subsection{${\cal P}(S^1)$ comme espace affine; d\'eriv\'ee de Schwarz}
${\cal P}(S^1)$ est un espace affine attach\'e \`a l'espace vectoriel ${\cal
L}^2$ des op\'erateurs de Hill. C'est-\`a-dire que la donn\'ee de 2 structures
$\alpha $ et $\beta \in {\cal P}(S^1)$ d\'efinit un op\'erateur de Hill $L$. On
notera~:
$$L = \alpha  - \beta \eqno{(3)}$$

Cette construction appartient \`a Hubbard [H] (voir aussi [G] [I]).
\vskip0.3cm
{\bf D\'efinition~: D\'eriv\'ee de Schwarz.}
Soient $\varphi $ et $\psi $ deux immersions $I\!\!R \rightarrow I\!\!R P^1$.
Pour
tout $x\in I\!\!R$, il existe une unique homographie $h_x \in PGL(2\;; I\!\!R)$
telle
que les jets d'ordre 2 en $x$ des immersions $\varphi $ et $h\circ \psi $
co\"{\i}ncident. La d\'eriv\'ee d'ordre 3~:
$$(\varphi  - h_x \circ \psi )'''(x)$$
definit alors une application cubique $T_x I\!\!R \rightarrow T_{\varphi (x)}
I\!\!R
P^1$ et sa composition avec $(T_x\varphi )^{-1}\;: T_{\varphi (x)} I\!\!R P^1
\rightarrow T_x I\!\!R$ est une application cubique $T_x I\!\!R \rightarrow T_x
I\!\!R$. Cette derni\`ere s'identifie \`a une forme quadratique sur $T_x
I\!\!R$ qui
sera not\'ee $S(\varphi ,\psi )(x)$. La diff\'erentielle quadratique $S(\varphi
,\psi
)$ est appel\'ee {\it d\'eriv\'ee de Schwarz} (g\'en\'eralis\'ee) de $\varphi $
par
rapport \`a $\psi $.
\vskip0.3cm
En particulier si $\psi (x) = x$ (modulo l'identification $I\!\!R P^1\cong
I\!\!R\cup\{\infty\}$) on obtient la d\'eriv\'ee schwarzienne classique~:
$$S(\varphi ) = \left[\frac{\varphi'''}{\varphi '} -
\frac{3}{2}\left(\frac{\varphi''}{\varphi '}\right)^2\right] (dx)^2\;\;\;\;
\eqno{(4)}$$
les deux propri\'et\'es fondamentales de la d\'eriv\'ee de Schwarz
sont~:

\vskip0.3cm
(j) $S(h\circ\varphi ,\psi ) = S(\varphi , h\circ \psi ) = S(\varphi ,\psi
)\;\;\;\;
\forall h\in PGL(2\;; I\!\!R)$
(invariance projective)

(jj) $S(\varphi ,\psi ) + S(\psi ,\chi) = S(\varphi ,\chi)\;\;\;\;(\chi\in
\mbox{Imm}^\infty(I\!\!R\;;\;I\!\!R P^1))$
\vskip0.3cm

consid\'erons \`a pr\'esent $\varphi $ et $\psi $ comme \'etant deux
applications
d\'eveloppantes pour les deux structures projectives $\alpha $ et $\beta \in
{\cal
P}(S^1)$. La diff\'erentielle quadratique $S(\varphi ,\psi )$ ne d\'epend que
de la
donn\'ee de $\alpha $ et $\beta $ (propri\'et\'e (i)).
Elle de plus d\'efinie sur $S^1$. On d\'efinit alors~:

$$\alpha - \beta \;: = S(\varphi , \psi )$$

L'op\'erateur de Hill $(3)$ admet pour potentiel $u(x) = S(\varphi ,\psi
)(x)$ (par construction).

Les deux propri\'et\'es suivantes sont alors v\'erifi\'ees~:
\vskip0.3cm
(a) $\forall\alpha ,\beta ,\gamma \in {\cal P}(S^1)\;\;,\;\;(\alpha -\beta )+
(\beta
-\gamma) = \alpha -\gamma\;$ (relation de Chasles)
\vskip0.3cm
(b) Pour toute structure $\alpha \in {\cal P}(S^1)$ et toute diff\'erentielle
quadratique $q$ sur $S^1$, il existe une unique structure $\beta \in {\cal
P}(S^1)$
telle que $q = \alpha -\beta $.
\vskip 0.3cm
Ces deux derni\`eres propri\'et\'es ach\`event de d\'emontrer le fait que
${\cal
P}(S^1)$ est un espace affine attach\'e \`a ${\cal L}^2$.

\subsection{Action du groupe des diff\'eomorphismes}

Soit $\Diff(I\!\!R)$ le groupe des diff\'eomorphismes de classe $C^\infty$ de
la
droite r\'eelle. Il agit sur l'espace des courbes param\'etr\'ees $\Gamma(M)$
d'une
vari\'et\'e $M$ par reparam\'etrisations. L'isomorphisme (2) permet de
d\'efinir une action de $\Diff(I\!\!R)$ sur l'espace ${\cal L}^n$ des
op\'erateurs
d'ordre $n$.

\begin{lemme}
(i) Le potentiel de degr\'e $n-2$ dans un op\'erateur $L\in {\cal L}^n$ se
transforme
sous l'action d'un diff\'eomorphisme $g\in \Diff (I\!\!R)$ de la fa\c con
suivante~:
$$g^\ast u_{n-2} = (u_{n-2}\circ g) . g'^2 + \frac{n(n-1)(n+1)}{12} S(g)$$

(ii) L'expression~:
$$w_3 = u_{n-3} - \frac{n-2}{2} \times u'_{n-2}$$
se transforme comme une diff\'erentielle cubique~:
$$g^\ast w_3 = (w_3 \circ g) \times g'^3$$
\end{lemme}

\noindent{\bf Preuve~:} il s'agit ici d'un calcul direct.$\Box$
\vskip0.4cm
Les autres coefficients $u_i$ se transforment, sous l'action de
$\Diff(I\!\!R)$,
suivant des formules plus compliqu\'ees. Cependant, nous avons tout de m\^eme
le fait
classique suivant~:

\begin{proposition}Il existe une s\'erie de polyn\^omes diff\'erentiels en
$u_i$~:
$$w_k = w_k (u_i, u'_i,\ldots)\;\;\;k\in \{3,4,\ldots,n\}$$
qui se transforment pour l'action de $\Diff(I\!\!R)$ comme des
diff\'erentielles de
degr\'e $k$~:
$$w_k = w_k(x) (dx)^k$$
\end{proposition}

Les expressions (non explicites) pour les $w_k$ ont \'et\'e d\'etermin\'ees
pour
la premi\`ere fois dans la litt\'erature physique (voir [F-I-Z]).
\vskip0.4cm
\noindent{\bf Remarque~:} Les diff\'erentielles $w_k$ ne sont pas uniquement
d\'efinies pour $k > 3$.
\vskip 0.3cm
Il est \'egalement possible d'interpr\'eter g\'eom\'etriquement les
solutions de l'\'equation diff\'erentielle lin\'eaire $Ly = 0$ comme le montre
le
lemme suivant.

\begin{lemme} Sous l'action du groupe $\Diff(I\!\!R)$, les solutions de
l'\'equation
diff\'erentielle $Ly =0$ se transforment comme des densit\'es tensorielles de
poids
$\frac{1-n}{2}$~:
$$y = y(x) (dx)^{\frac{1-n}{2}}$$
\end{lemme}

\noindent{\bf Preuve~:} (i) Rappelons que l'action de $\Diff(I\!\!R)$ sur
${\cal L}^n$
est donn\'ee par son action sur les courbes $\gamma\subset I\!\!R P^{n-1}$.
Nous
allons \`a pr\'esent expliciter les formules donnant les solutions de
l'\'equation
diff\'erentielle $L_\gamma y =0$.

Soit $(\gamma_1(x),\ldots, \gamma_{n-1}(x))$
un syst\`eme de coordonn\'ees locales affines pour la courbe $\gamma$. Notons
$\overline
w_\gamma$ la fonction d\'efinie par~:
$$\overline w_\gamma(x) = \left\vert \matrix{\gamma _1' \hfill &\dots & \gamma
_{n-1}'\hfill\cr
&\dots\hfill\cr
\gamma _1^{(n-1)} \hfill &\dots &\gamma _{n-1}^{(n-1)}}\right\vert $$
Les $n$ fonctions:
$$y_0 = \overline w_\gamma^{-\frac{1}{n}},
y_1=\gamma _1\overline w_\gamma^{-\frac{1}{n}},\dots ,y_{n-1}=
\gamma _{n-1}\overline w_\gamma^{-\frac{1}{n}}\eqno{(5)}$$
constituent alors un syst\`eme fondamental de solution pour l'\'equation
$$L_\gamma y
= 0$$

(ii) La formule (5)  montre imm\'ediatement que les solutions $y_i$ se
transforment sous l'action des reparam\'etrisations $g^\ast\gamma = \gamma\circ
g$ de
la fa\c con suivante~:
$$g^\ast y = (y\circ g) \times (g')^{\frac{1-n}{2}}\hskip2cm\Box$$
\vskip 0.3cm
{\bf Exemples~:}

($n=2$)  Les solutions d'une \'equation de Sturm-Liouville se transforment
comme
des densit\'es tensorielles de poids $-1/2$ sur $I\!\!R$.
\vskip0.3cm
($n=3$)  Les solutions de l'\'equation diff\'erentielle $y''' + u_1(x) y' +
u_0(x) y
= 0$ se transforment comme des champs de vecteurs sur $I\!\!R$.
\vskip0.3cm
{\bf Monodromie.~}
Soient $\gamma\in \Gamma_T(I\!\!R P^{n-1})$ une courbe T-\'equivariante.
L'action d'un
diff\'eomorphisme $g\in \Diff^+(I\!\!R)$ pr\'eserve la monodromie si et
seulement si~:
$$g(x + 2\pi) = g(x) + 2\pi\hskip2cm \forall x\in I\!\!R$$

Ces diff\'eomorphismes constituent un sous-groupe de $\Diff^+(I\!\!R)$
isomorphe au
rev\^etement universel du groupe $\Diff^+(\S1)$ des diff\'eomorphismes du
cercle
respectant l'orientation. Par cons\'equent, l'action de
$\widetilde{\Diff^+(\S1)}$ sur
les courbes \'equivariantes se projette en une action de $\Diff^+(\S1)$ sur
l'espace
${\cal L}^n_{2\pi}$ des op\'erateurs diff\'erentiels \`a {\it coefficients
p\'eriodiques}.

Il est \`a remarquer que -- en ce qui concerne l'action des diff\'eomorphismes
sur
les solutions -- c'est {\it toujours} $\widetilde{\Diff^+(\S1)}$ qui agit et
non
$\Diff^+(\S1)$.
\vskip0.3cm
{\bf Exemple: l'espace $\overline\Gamma_0(I\!\!R P^2)$.~}
Consid\'erons l'espace $\overline\Gamma_0(I\!\!R P^2)$ de toutes les courbes
param\'etr\'ees, non d\'eg\'en\'er\'ees et \'equivariantes modulo \'equivalence
projective. Notons ${\cal F}_3(S^1)$ l'espace des diff\'erentielles cubiques
sur
$S^1$ et $q\;: {\cal P}(S^1)\rightarrow {\cal P}(S^1)\times {\cal F}_3(S^1)$ le
plongement~: $q(\alpha )=(\alpha ,0)$.
\vskip0.3cm
Il existe alors un isomorphisme $\Phi\;: \overline\Gamma_0(I\!\!R
P^2)\rightarrow
{\cal P}(S^1)\times {\cal F}_3(S^1)$ rendant coh\'erent le diagramme suivant :
$$\diagram{
{\cal L}^3_{2\pi}\cong\overline\Gamma_0(I\!\!R P^2)&\displaystyle\hfl{\Phi}{}&
{\cal P}(S^1)\times {\cal F}_3(S^1)\cr
\noalign{\vskip 7mm}
\displaystyle \mu\searrow{}{}&&
\nearrow{\displaystyle q}{}\cr
\noalign{\vskip 4mm}
&{\cal P}(S^1)\cong {\cal L}^2_{2\pi}&\cr}$$
o\`u $\mu$ d\'esigne l'application $\Diff^+(\S1)$ -- \'equivariante~:
$$\partial^3 + u_1\partial + u_0 \stackrel{\mu}{\longmapsto}\partial^2 +
\frac{u_1}{4}$$

La construction de l'application $\Phi$ est la suivante~:
Soit $L = \partial^3 + u_1 \partial + u_0$. Alors,
$$\Phi(L) = (\mu(L)\;; w_3) = \left(\partial^2 + \frac{u_1}{4}\;;(u_0
-\frac{u'_1}{2} (dx)^3\right)$$

Remarquons que $(\Phi^{-1}\circ q) (\partial^2 + u) = \partial^3 + 4u \partial
+ u'$
et que la courbe $\overline\gamma\in \overline\Gamma_0(I\!\!R P^2)$ associ\'ee
\`a
cet op\'erateur a son support dans une c\^onique de $I\!\!R P^2$ (voir [C]).

\subsection{Longueur d'arc affine et projective~; courbure affine et
projective.}

{\bf Longueur d'arc affine.~}
Soit $c\in {\cal C}(I\!\!R^n)$ une courbe non param\'etr\'ee et non
d\'eg\'en\'er\'ee. Fixons un param\`etre $x$ quelconque sur $c$.

\vskip0.3cm
{\bf D\'efinition~:}
On appelle {\it longueur d'arc affine} sur $c$ la 1-forme
$$d\sigma  = C^{te}\left[\overline W_c(x)\right]^{\frac{2}{n(n+1)}}dx$$
o\`u $\overline W_c(x) = \mbox{det}(c'(x), c''(x), \ldots, c^{(n)} (x))$
\vskip0.3cm
Le param\`etre $\sigma $ est d\'efini \`a une transformation affine pr\`es~:
$\sigma
\sim a\sigma  + b\;(a\in I\!\!R\backslash\{0\}, b\in I\!\!R).\;\;\sigma $ est
appel\'e le {\it param\`etre affine} sur $c$.

\vskip0.3cm
Il est \`a noter que la d\'efinition du param\`etre affine fait intervenir le
volume
euclidien du parall\'el\'epip\`ede support\'e par les $n$ vecteurs
$c'(x),\ldots,c^{(n)}(x)$.
\vskip0.3cm
{\bf Courbure affine -- D\'efinition~:}
Les $n-1$ fonctions~:
$$\kappa_i(\sigma ) = (-1)^i det (c',\ldots, \hat c_i,\ldots, c^{(n+1)})(\sigma
)$$
(o\`u $ i\in
\{1,\ldots,n-1\}$ et le symbole R$\hat{\blanc}$S signifie l'ommission du
vecteur correspondant)
sur la courbe $c$ sont appel\'ees les {\it courbures affines} d'ordre $i$
de $c$.
\vskip0.3cm

{\bf Exemple~:}
Nous retrouvons dans le cas o\`u $n=2$ les d\'efinitions classiques de la
longueur
d'arc et de la courbure affine d'une courbe de $I\!\!R^2$~:
$$d\sigma=\left|\begin{array}{cc}
c'_1(x)  & c'_2(x)\\
\noalign{\vskip 3mm}
c''_1(x) & c''_2(x)\end{array}\right|^\frac{1}{3} dx \;\;\mbox{et} \;\;
\kappa (\sigma)=\left|\begin{array}{cc}
c''_1(\sigma)&c''_2(\sigma)\\
\noalign{\vskip 3mm}
c'''_1(\sigma) & c'''_2(\sigma)\end{array}\right|$$
si $\displaystyle c(x)=\left(\begin{array}{c}
c_1(x)\\
c_2(x)\end{array}\right)$. (voir [Bo] et [Bu]).
\vskip0.3cm
On peut en fait r\'ealiser les courbures $\kappa_i$ comme les
coefficients d'un op\'erateur diff\'erentiel lin\'eaire de degr\'e
$n+1$. Plus pr\'ecis\'ement, les coordonn\'ees de la courbe $c$
param\'etr\'ee par $\sigma$ v\'erifient l'\'equation
diff\'erentielle~:

$$y^{(n+1)}+ \kappa_1(\sigma) y^{(n-1)}+\ldots + \kappa_{n-1}(\sigma) y' =
0$$

{\bf Application de Gauss}.
Il est possible d'associer naturellement \`a chaque courbe affine
$c\in {\cal C}(I\!\!R^n)$ une courbe projective param\'etr\'ee
$\tau (c)\in \overline\Gamma(I\!\!R P^{n-1})$ modulo \'equivalence
projective.
\vskip0.3cm
Pr\'ecisons cette construction.
Munissons la courbe $c$ du param\`etre affine $\sigma $. La vitesse
$\dot c={d\over d\sigma }c$:

$$\dot c\;: I\!\!R\rightarrow I\!\!R P^{n-1}$$
$\tau(c)$ est alors la classe de $\dot c$ modulo $PGL(n\;; I\!\!R)$.

\begin{lemme} L'op\'erateur diff\'erentiel lin\'eaire
associ\'e \`a $\tau (c)\in  \overline\Gamma(I\!\!R P^{n-1})$ est~:
$$L_c = \partial^n_\sigma  + K_1(\sigma ) \partial^{n-2}_\sigma
+\ldots + K_{n-1}(\sigma )\;\;\;\; (\partial_\sigma
=\frac{d}{d\sigma })$$
\end{lemme}

\noindent{\bf Preuve~:} D'apr\`es (6) les coordonn\'ees
$y_i$ de $c'(\sigma )$ v\'erifient l'\'equation
$$L_c . y_i =0$$

On conclut en remarquant que $\tau (c)$ est la projectivis\'ee de
$(y_1,\ldots,y_n)\in I\!\!R^n$ via la projection
$\displaystyle I\!\!R^n \backslash\{0\}\rightarrow I\!\!R
P^{n-1}.\;\;\Box$
\vskip0.5cm
{\bf Longueur d'arc projective~}.
Soit $c\in {\cal C}(I\!\!R P^{n-1})$ une courbe non param\'etr\'ee et
non d\'eg\'en\'er\'ee dans l'espace projectif de dimension $n-1$ et
notons $x$ un param\`etre quelconque sur cette courbe .
Rappelons qu'\`a la courbe param\'etr\'ee $c(x)$ correspond un
op\'erateur diff\'erentiel $L_c\in {\cal L}^n$.

\vskip0.3cm
{\bf D\'efinition~:}

(i) On appelle {\it longueur d'arc projective} sur $c$ la 1-forme~:
$$d\sigma =[w_3(x)]^{\frac{1}{3}} dx$$
o\`u $w_3(x)=\left(u_{n-3}-\frac{1}{n-1} u'_{n-2}\right)(x) (dx)^3$
est la diff\'erentielle cubique associ\'ee \`a l'op\'erateur $L_c$.
\vskip0.4cm
(ii) Supposons que $w_3\not= 0$. Dans ce cas, $\sigma $ d\'efinit un
param\`etre local sur la courbe $c$ appel\'e {\it param\`etre
projectif}. Ces points sont caract\'eris\'es par le fait que la c\^onique
osculatrice poss\`ede un ordre de contact (en ce point) $\geq 5$ (voir
[C]).
\vskip0.3cm
{\bf Courbure projective~}
La diff\'erentielle quadratique $u_{n-2}(x)(dx)^2$ d\'efinit une
structure projective sur la courbe $c$. Plus pr\'ecis\'ement, il
existe une projection $\Diff (I\!\!R)$- \'equivariante de l'espace
${\cal L}^n$ sur l'espace ${\cal L}^2$ des op\'erateurs de
Sturm-Liouville~:
$$\mu\;: {\cal L}^n \rightarrow {\cal L}^2\;: L
\longmapsto\frac{n(n-1)(n+1)}{6} . \partial^2 +
u_{n-2}(x)\eqno{(6)}$$

{\bf D\'efinition~:}
On appelle {\it courbure projective} d'une courbe $c\in {\cal C}(I\!\!R
P^{n-1})$ la structure projective sur $c$ d\'efinie par
l'application $\mu$.
\vskip0.3cm

{\bf Exemple~:$(n=3)$~}
Au voisinage d'un point non sextactique, une courbe $c\in {\cal
C}(I\!\!R P^2)$ est uniquement d\'efinie (modulo $PGL(3\;; I\!\!R)$)
par sa courbure projective (cf [C] et Chap.5.6.).
\vskip0.3cm
{\bf Le th\'eor\`eme des six sommets~}
Les deux r\'esultats classiques suivants pour les courbes planes
ferm\'ees sont des analogue du th\'eor\`eme des quatre sommets de la
g\'eom\'etrie diff\'erentielle euclidienne plane.

\begin{lemme} Toute courbe ferm\'ee et non
d\'eg\'en\'er\'ee dans $I\!\!R P^2$ poss\`ede au moins 6 points
sextactiques.
\end{lemme}

\noindent{\bf Preuve~:} (voir par exemple [Bo] ou [Bu])
Ce r\'esultat d\'ecoule facilement du fait suivant~: fixons un
param\`etre quelconque $x$ sur notre courbe $c\in {\cal
C}_{id}(I\!\!R P^2)$ et notons $L$ l'op\'erateur diff\'erentiel de
degr\'e 3 associ\'e \`a la courbe param\'etr\'ee $c(x)$. Pour toute
solution $y$ de l'\'equation diff\'erentielle $Ly =0$, on a~:
$$\int_{S^1}w_3 (x) y(x)^2 dx = 0\;\;\;\Box$$
Convenons d'appeler {\it sommet} d'une courbe affine de $I\!\!R^2$
tout point critique pour sa courbure. On a alors le~:

\begin{corollaire} Une courbe affine ferm\'ee et non
d\'eg\'en\'er\'ee dans $I\!\!R^2$ poss\`ede au moins 6 sommets.
\end{corollaire}

\noindent{\bf Preuve~:} Soit $c\in {\cal C}_{id}(I\!\!R^2)$
et munissons cette courbe de son param\`etre affine $\sigma $.
\vskip0.5cm
Ses  coordonn\'ees $c_1(\sigma )$ et $c_2(\sigma )$ v\'erifient
alors l'\'equation~:
$$c'''_i(\sigma ) + \kappa(\sigma ) c'_i(\sigma )=0$$
A l'op\'erateur diff\'erentiel $\partial^3 + \kappa(\sigma )\partial$
correspond une classe de courbes ferm\'ees et non d\'eg\'en\'er\'ees
$\overline\gamma\in\overline\Gamma_{id}(I\!\!R P^2)$. D'apr\`es le
lemme pr\'ec\'edent, chacune de ces courbes poss\`ede au moins 6
points sextactiques et, par cons\'equent, la diff\'erentielle
cubique $\displaystyle w_3(\sigma )=-\frac{1}{2}\kappa'(\sigma )$
s'annule au moins en 6 points.$\;\;\Box$
\vskip0.3cm
{Cas des courbes planes \'equivariantes}

\begin{lemme} Si l'op\'erateur de monodromie $T$ d'une courbe
\'equivariante $c\in {\cal C}_T(I\!\!R P^2)$ poss\`ede une de ses valeurs
propres \'egale \`a l'unit\'e, alors la courbe $c$ a au moins 2 points
sextactiques.
\end{lemme}

\begin{corollaire} Une courbe affine \'equivariante $c\in {\cal
C}_0(I\!\!R^2)$ poss\`ede au moins 2 sommets.
\end{corollaire}

Les preuves de ces deux r\'esultats sont analogues \`a celles de
2.6.6.

\subsection{L'espace ${\cal P}_T(S^1)$ comme vari\'et\'e
symplectique}

L'espace ${\cal P}(S^1)$ form\'e par toutes les structures
projectives sur le cercle unit\'e $S^1$ est munie d'une structure de
Poisson [Ki1].

Sur chaque sous-espace ${\cal P}_T(S^1)$ constitu\'e des structures
projectives \`a monodromie $T$ fix\'ee, ce crochet de Poisson est
sous-jacent \`a une structure symplectique sur ${\cal
P}_T(S^1)\;\;:\;{\cal P}_T(S^1)$ est une ``feuille symplectique'' de
ce crochet (voir [Se1]).

Nous pr\'esentons ici la formule explicite pour la forme
symplectique de l'espace ${\cal P}_T(S^1)$. La d\'efinition du
crochet de Poisson sera donn\'ee (dans le cas le plus g\'en\'eral)
au chapitre 7.

\begin{lemme} Soit $\alpha \in {\cal P}_T(S^1)$, on a alors~:
$$T_\alpha {\cal P}_T(S^1) \cong \mbox{Vect}(S^1)\biggm/aut(\alpha
)$$
o\`u Vect$(S^1)$ d\'esigne l'alg\`ebre de Lie des champs de vecteurs
de classe $C^\infty$ sur le cercle et $aut(\alpha )\hookrightarrow
Vect(S^1)$ est l'alg\`ebre de Lie du groupe des automorphismes de la
structure projective $\alpha $.
\end{lemme}

\noindent{\bf Preuve~:} L'espace ${\cal P}_T(S^1)$ est un espace
homog\`ene du groupe $\Diff^+(\S1)$, ce dernier agissant par
transport de structure sur ${\cal P}(S^1).\;\;\;\Box$
\vskip0.5cm

Fixons une fois pour toutes une forme longueur $dt$ sur la droite
projective $I\!\!R P^1$. La donn\'ee d'une application
d\'eveloppante $\varphi \;: I\!\!R\rightarrow I\!\!R P^1$ pour une
structure $\alpha \in {\cal P}_T(S^1)$ est \'equivalente \`a la
donn\'ee de la 1-forme $\theta =\varphi ^\ast(dt)$. Si $\delta\alpha$
d\'esigne un vecteur tangent \`a ${\cal P}_T(S^1)$ en $\alpha $, il
existe un champ $v\in Vect(S^1)$ tel que~: $\delta \alpha = L_v
\theta $ (o\`u $L_v$ d\'esigne la d\'eriv\'ee de Lie par rapport au
champ $v$). On peut en outre choisir ce champ de fa\c con \`a ce
qu'il s'annule en un point.$\;\;\;\Box$
\vskip 0.5cm
{\bf D\'efinition de la forme symplectique~}
On d\'efinit une 2-forme symplectique $\omega $ sur la vari\'et\'e
${\cal P}_T(S^1)$ de la fa\c con suivante. Soit $\delta_1\alpha $ et
$\delta _2\alpha \in T_\alpha {\cal P}_T(S^1)$ deux vecteurs
tangents en la structure $\alpha \in {\cal P}_T(S^1)$. Soit $\xi _1$
et $\xi _2\in Vect(S^1)$ deux champs sur $S^1$ tels que~: $\delta
_1\alpha = L_{\xi _i}\theta $ et $\xi _i(x_0)=0$ pour un point
$x_0\in S^1$, fix\'e.

On pose alors~:
$$\omega (\delta _1\alpha , \delta _2\alpha )=\int_{S^1}\frac{L_{\xi
_1}\theta }{\theta }d \frac{L_{\xi _2}\theta }{\theta
}\eqno{(7)}$$

Remarquons que $L_{\xi _i}\theta $ est une 1-forme sur $I\!\!R$, mais
que la fonction $\dsp{L_{\xi _i}\theta \over \theta}$ est
$2\pi$-p\'eriodique, donc d\'efinie sur $S^1$.
\vskip0.5cm

\begin{lemme} La 2-forme $\omega $ est bien d\'efinie sur ${\cal
P}_T(S^1)$~; elle est ferm\'ee, non d\'eg\'en\'er\'ee et sa d\'efinition
est ind\'ependante du choix de $\xi _1, \xi _2,\varphi $ et du
param\`etre $t$.
\end{lemme}

\noindent{\bf Preuve~:} En utilisant le param\`etre $t$, on a les
\'ecritures suivantes pour la 1-forme $\theta $ et les champs $\xi
_i$~:
$$\theta =\varphi '(t)dt\;\;\;\;\xi _i = \xi _i(t)
\frac{d}{dt}\;\;\;\xi _i(t + 2\pi) = \xi _i(t)$$

Un calcul direct nous donne acc\`es \`a la formule suivante~:
$$\omega (\delta _1\alpha , \delta _2\alpha )= \int_{S^1}S(\varphi
)(t)[\xi _1, \xi _2](t) dt + \int_{S^1}\xi '_1(t) \xi ''_2(t)dt$$
o\`u $S(\varphi )$ est la d\'eriv\'ee de Schwarz (classique) et
$[\xi _1, \xi _2]=\xi _1\;\xi'_2 - \xi '_1\;\xi _2$ d\'esigne le
crochet de Lie des champs de vecteurs sur $S^1$.

L'invariance de la formule (7) vis-\`a-vis du choix de $\xi _1,
\xi _2, \varphi $ et $dt$ peut \^etre v\'erifi\'ee soit directement,
soit plus rapidement en utilisant la~:
\vskip0.3cm
{\bf Remarque\footnote{Voir aussi 2.7.5.}~:} La formule
(7) co\"{\i}ncide avec la forme symplectique de
Kirillov-Kostant-Souriau sur une orbite coadjointe de l'alg\`ebre de
Virasoro (voir [Ki1]).$\;\;\;\Box$
\vskip0.5cm

\begin{lemme}
La forme $\omega $ est invariante relativement \`a l'action du
groupe $\Diff^+(\S1)$ sur l'espace ${\cal P}_T(S^1)$.
\end{lemme}

{\bf Alg\`ebre de Virasoro~: crochet de Poisson sur
l'espace ${\cal P}(S^1)$~}
Consid\'erons l'action du groupe $\Diff^+(\S1)$ sur l'espace
${\cal L}^2_{2\pi}$ des op\'erateurs de Hill $\partial^2 + u(x)$ (cf
2.5.1). L'action infinit\'esimale associ\'ee est une action de l'alg\`ebre
de Lie Vect$(S^1)$ sur ${\cal L}^2_{2\pi}$~:
$$L_f u = f u' + 2f' u + \frac{f'''}{2}$$
o\`u $f =f(x)\frac{d}{dx} \in Vect(S^1)$ et $L_f$ d\'esigne
la d\'eriv\'ee de Lie.
\vskip0.5cm

{\bf D\'efinition~:}
Soit
$$\ell_f\;: {\cal L}^2_{2\pi}\rightarrow I\!\!R\;:\;
u\longmapsto \int_{S^1}f(x) u(x) dx$$
une fonctionnelle
lin\'eaire sur l'espace des op\'erateurs de Hill. On lui associe un
champ de vecteurs $\xi $ sur ${\cal L}^2_{2\pi}$ en posant~:
$$\xi  = L_f$$

\begin{lemme} L'application $\xi \;: ({\cal
L}^2_{2\pi})^{\star }\rightarrow {\cal L}^2_{2\pi}$ est un op\'erateur
hamiltonien sur ${\cal L}^2_{2\pi}$: il d\'efinit un crochet de
Poisson sur ${\cal P}(S^1)$.
\end{lemme}

\noindent{\bf Preuve~:} En g\'en\'eral, pour d\'efinir un crochet de
Poisson sur une vari\'et\'e $V$, on commence par se donner un op\'erateur
$\xi \;: T^\ast V\rightarrow TV$ lin\'eaire et antisym\'etrique. Si $f$
et $g\in C^\infty(V)$, on pose~:
$$\{f,g\}\;:= < dg , \xi (df) > \in C^\infty(V)$$
$\{\;.,\;.\;\}$ d\'efinit un crochet de Poisson sur $V$ si~:
$$\sum_{cycl. }\{ f,\{g,h\}\}=0\hskip1cm\forall f , g, h \in
C^\infty(V)$$
c'est l'identit\'e de Jacobi.

Pour v\'erifier l'identit\'e de Jacobi dans notre cas, il suffit de la
v\'erifier sur les fonctionnelles $\ell _f$. Posons donc~:
$$\{\ell _f,\ell _g\}(u) = < \ell _g, Lg(u) > = \int_{S^1} g(f u' +
2f 'u + \frac{1}{2}f''')(x) dx$$
o\`u $f,g\in Vect(S^1)$ et $\partial^2 + u(x) \in {\cal L}^2_{2\pi}$.

On s'apper\c coit alors que ces fonctionnelles lin\'eaires d\'efinissent
une alg\`ebre de Lie~:
$$\{\ell _f,\ell_g\}(u) = \ell _{[f,g]}(u) + \int_{S^1}f '(x)
g''(x) dx\;\;\;\eqno{(8)}$$

\hskip9cm$\Box$
\vskip0.4cm
{\bf Remarque~:} L'alg\`ebre de Lie d\'efinie par (8)
s'appelle {\it l'alg\`ebre de Virasoro}. C'est une extension centrale de
l'alg\`ebre de Lie $Vect(S^1)$.
\vskip0.3cm
Le fait suivant peut \^etre d\'emontr\'e en utilisant la formule
(8):
\vskip0.3cm

\begin{proposition}
Une orbite coadjointe de l'alg\`ebre de Virasoro s'identifie \`a un
espace ${\cal P}_T(S^1)$. La forme symplectique sur ce dernier espace
co\"{\i}ncide alors avec la 2-forme symplectique canonique sur l'orbite
coadjointe.
\end{proposition}

\vfill\eject

\section{STRUCTURES SYMPLECTIQUES SUR L'ESPACE DES COURBES AFFINES}

Soit $M$ une vari\'et\'e localement affine (cf. Chap. 2.1) de
dimension $n$. Nous d\'efinissons ici une forme symplectique sur
l'espace $\overline{\cal C}_{id}(M)$ des courbes ferm\'ees, non
d\'eg\'en\'er\'ees et non param\'etr\'ees sur $M$ \`a \'equivalence
affine pr\`es.

Notre d\'efinition repose sur deux id\'ees principales~:
\vskip0.5cm
(a) A chaque courbe $c\in {\cal C}(M)$ est associ\'ee une classe
dans $\overline\Gamma(I\!\!R P^{n-1})$ (l'espace des courbes {\it
param\'etr\'ees} et non d\'eg\'en\'er\'ees sur $I\!\!R P^{n-1}$
modulo \'equivalence projective). Nous d\'efinissons alors une
application~:
$$\Pi\;: {\cal C}_{id}(M)\longrightarrow\overline\Gamma(I\!\!R
P^{n-1})$$
dans l'espace des courbes projectives param\'etr\'ees (modulo
\'equivalence projective). Cette application est affinement
invariante et induit une application~:
$$\overline\Pi\;: \overline{\cal C}_{id}(M)\longrightarrow
\overline\Gamma(I\!\!R P^{n-1})$$

Sur chaque composante connexe de l'espace $\overline{\cal
C}_{id}(M)$, l'image de $\overline\Pi$ est incluse dans le
sous-espace des courbes projectives \`a {\it monodromie fix\'ee}.
\vskip0.5cm
(b) Il existe une structure symplectique canonique sur chaque espace
$\overline\Gamma_T(I\!\!R P^{n-1})$ (dans le cas o\`u $n=2$, cette
structure a \'et\'e d\'efinie dans le chap. 2.7~; pour le cas o\`u
$n\geq 3$ voir chap. 6.2).

\subsection{Cas o\`u $dim(M)=2$.}

Pla\c cons-nous \`a pr\'esent dans le cas d'une {\it surface}
localement affine $M$.

Si $c\in {\cal C}(M)$, nous pouvons d\'efinir deux structures
g\'eom\'etriques sur $c$ (cf. Chap. 2.6)~:
\vskip0.5cm
(i) Une param\'etrisation affine~:
$$\sigma \;: I\!\!R \rightarrow c$$
\vskip0.5cm
(ii) Son application de Gauss~:
$$\tau \;: c\rightarrow I\!\!R P^1$$

La premi\`ere est une structure {\bf affine} sur $c$ et la seconde
une structure {\bf projective}.
\vskip 0.3cm
{\bf D\'efinition de l'application $\Pi\;:
{\cal C}_{id}(M)\rightarrow{\cal P}(S^1)$~}
On peut interpr\'eter $\sigma $ et $\tau $ comme deux structures
projectives sur la courbe ferm\'ee $c\approx S^1$ (pas de mani\`ere
canonique). L'espace ${\cal P}(S^1)$ est un espace affine model\'e
sur l'espace vectoriel des op\'erateurs de Hill (voir Chap. 2.4.3).
Fixons alors une structure projective $\alpha \in {\cal P}(S^1)$ qui
jouera le r\^ole de ``l'origine'' (on pourra prendre par exemple la
structure projective canonique d\'efinie par le rev\^etement
universel de $I\!\!R P^1$).
\vskip0.3cm
$\Pi(c)$ est alors d\'efinie par l'\'egalit\'e~:
$$\tau  - \sigma  = \Pi(c) - \alpha  \;\;\in \;{\cal L}^2_{2\pi}$$

Nous pouvons \'egalement donner une version plus na\''{\i}ve, quoique
directe, de cette d\'efinition~: on obtient une immersion $\pi_c\;:
I\!\!R \rightarrow I\!\!R P^1$ en posant~:
$$\pi_c := \tau \circ \sigma $$
Cette immersion $\pi_c$ est l'application d\'eveloppante d'une
unique structure projective sur $S^1$. Nous appellerons $\pi(c)$
cette derni\`ere.

\begin{lemme}L'application $\Pi$ ``descend'' sur l'espace quotient
$\overline{\cal C}_{id}(M)$.
\end{lemme}

\noindent{\bf Preuve~:} $\Pi$ est -- par construction -- affinement
invariante.$\;\;\Box$
\vskip0.3cm
On obtient donc une application $\overline\Pi\;: \overline{\cal
C}_{id}(M)\rightarrow{\cal P}(S^1)$.

\begin{lemme}$\overline\Pi$ envoit chaque composante connexe de $\overline{\cal
C}_{id}(M)$ dans un sous-espace ${\cal P}_T(S^1)$.
\end{lemme}

\noindent{\bf Preuve~:} Faisons le choix d'une paire d\'eveloppante
$(\varphi ,T)$ pour la structure affine de la surface $M$~:
$$\left\{\begin{array}{ll}
\varphi \;:\widetilde M\rightarrow I\!\!R^2\\
\noalign{\vskip 3mm}
T\;: \pi_1(M) \rightarrow Aff(2\;; I\!\!R)\end{array}\right.$$

Soit $c\in {\cal C}_{id}(M)$ une courbe ferm\'ee sur la surface $M$
et $\widetilde c\subset \widetilde M$ un relev\'e quelconque de
cette courbe dans le rev\^etement universel de la surface. La classe
d'homotopie [c] de notre courbe peut \^etre consid\'er\'ee comme un
\'el\'ement de $\pi_1(M)$. Posons donc $A\;:= T([c])\in Aff(2\;;
I\!\!R)$. La courbe $\varphi (\widetilde c)\subset I\!\!R^2$ est
A-\'equivariante~:
$$\varphi (\widetilde c) \in {\cal C}_A(I\!\!R^2)$$

Notons $T\in PGL(2\;; I\!\!R)$ l'image de $A$ par la projection
naturelle~:
$$Aff(2\;; I\!\!R)\rightarrow PGL(2\;; I\!\!R)$$

Munissons $c$ de son param\`etre affine $\sigma $ et soit $\beta \;:
I\!\!R \rightarrow I\!\!R P^1$ l'application de Gauss associ\'ee \`a
la courbe $(\varphi \circ\widetilde c)(\sigma )$. On obtient~:
$$\left\{\begin{array}{ll}
\beta (\sigma +2\pi) = T. \beta (\sigma )\\
\noalign{\vskip 3mm}
\beta \hbox{ est une application d\'eveloppante pour la structure\ \ }
\Pi(c)\end{array}\right.$$

De plus, si $c$ reste dans la m\^eme classe d'homotopie, $A$ et, par
cons\'equent, $T$ restent fixes.$\;\;\Box$

\begin{proposition}
Sur chaque composante connexe $C$ de l'espace $\overline{\cal
C}_{id}(M)$, la restriction $\overline\Pi\;: C \rightarrow{\cal
P}_T(S^1)$ est une bijection.
\end{proposition}

\noindent{\bf Preuve~:} L'op\'erateur de Hill $L =\tau  - \sigma $
associ\'e \`a la structure projective $\Pi(c)\;\;(c\in {\cal
C}_{id}(M))$ est donn\'e par~:
$$L = \partial^2_\sigma  + \kappa(\sigma )$$
o\`u $\kappa(\sigma )$ d\'esigne les courbure affine de la courbe $c$
(voir Chap. 2.6).

Or, une courbe affine est d\'efinie uniquement par sa courbure,
modulo \'equivalence affine.$\;\;\Box$

\begin{corollaire}
Il existe sur chaque composante connexe de l'espace
$\overline{\cal C}_{id}(M)$ une structure de vari\'et\'e symplectique. La
forme symplectique $\Omega $ est donn\'ee par la formule~:
$$\Omega  = \overline \Pi^\ast(\omega )$$
$\omega $ d\'esignant la forme symplectique sur l'espace ${\cal
P}_T(S^1)$.
\end{corollaire}

\subsection{Cas o\`u $dim M = n \geq  3$  -- Fin de la d\'emonstration
du th\'eor\`eme 1.}

$M$ d\'esignera une vari\'et\'e affine de dimension $n\geq 3$.

Nous g\'en\'eralisons dans cette section la construction vue en 3.1.
l'existence d'une param\'etrisation affine pour une courbe $c\in
{\cal C}_{id}(M)$ conduit \`a l'existence d'un plongement~:
$${\cal C}_{id}(M) \hookrightarrow \Gamma (M)$$

De plus, le choix d'une application d\'eveloppante $\varphi
\;:\widetilde M\rightarrow I\!\!R^n$ entra\^{\i}ne l'existence d'une
application~:
$$\Gamma(M) \rightarrow \Gamma(I\!\!R P^{n-1})$$
(Application de Gauss associ\'ee \`a une courbe param\'etr\'ee).
Nous obtenons donc une application~:
$$\Pi\;: {\cal C}_{id}(M) \rightarrow \Gamma(I\!\!R P^{n-1})$$

Soit \`a pr\'esent $p\;: Aff(n\;;I\!\!R)\rightarrow PGL(n\;;
I\!\!R)$ la projection naturelle. On a le

\begin{lemme}
Soit $A\in Aff(n\;;I\!\!R)$ et $T\;:= p(A)$, on a~:
$$\Pi({\cal C}_A(I\!\!R^n))\subset \Gamma_T (I\!\!R P^{n-1})$$
\end{lemme}

La d\'emonstration est analogue \`a celle du lemme 3.1.3.
\vskip 0.3cm
Les autres \'etapes qui suivent ce lemme r\'ep\`etent le cas de la
dimension 2. Nous obtenons finalement l'application~:
$$\overline{\Pi}\colon\overline{\cal C}_{id}(M)
\rightarrow\overline\Gamma(I\!\!R P^{n-1})$$
Sur chaque composante connexe de l'espace
$\overline{\cal C}_{id}(M), \;\overline{\Pi}$ est une bijection sur un
sous-espace $\overline\Gamma_T(I\!\!R P^{n-1})$.

$\overline\Gamma_T(I\!\!R P^{n-1})$ est muni d'une structure de
vari\'et\'e symplectique (voir Chap. 6.2).

Nous d\'efinissons donc sur $\overline{\cal C}_{id}(M)$ une forme
symplectique, image r\'eciproque par $\overline\Pi$ de celle sur
$\overline\Gamma_T(I\!\!R P^{n-1})$. Le th\'eor\`eme 1 est
d\'emontr\'e.$\;\;\;\Box$
\vskip 0.3cm
{\bf Remarque~: Action du groupe $\Diff^+(\S1)$.~}
Le groupe $\Diff^+(\S1)$ agit sur l'espace $\overline\Gamma_T(I\!\!R
P^{n-1})$ par reparam\'etrisation des courbes. L'isomorphisme
$\overline\Pi$ nous permet de transporter cette action sur chaque
composante connexe de l'espace $\overline{\cal C}_{id}(M)$. Par
exemple, dans le cas de la dimension 2, le groupe $\Diff^+(\S1)$ agit
sur la courbure $\kappa(\sigma )$ d'une courbe $c\in {\cal C}_{id}(M)$
comme sur le potentiel de l'op\'erateur de Hill $\partial^2_\sigma
+ \kappa(\sigma )$.
\vfill\eject

\section{STRUCTURE DE POISSON SUR L'ESPACE DES COURBES DE CARTAN.}

Nous consid\'erons dans ce chapitre l'espace ${\cal C}_0(I\!\!R P^2)$
des courbes projectives \'equivariantes, non param\'etr\'ees et non
d\'eg\'en\'er\'ees dans le plan projectif.

La m\^eme id\'ee qui nous avait conduits \`a la construction d'une
structure symplectique sur l'espace des courbes affines, nous
am\`ene, dans le cas projectif, \`a la construction d'une structure
de Poisson. De la m\^eme mani\`ere que dans le chapitre 3, nous
allons associer \`a toute courbe $c\in {\cal C}_0(I\!\!R P^2)$ deux
structures g\'eom\'etriques sur cette courbe~:
\vskip0.5cm
(a) Une structure {\it affine}, engendr\'ee par la largeur d'arc
projective sur $c$.
\vskip0.5cm
(b) Une structure {\it projective}, engendr\'ee par la courbure
projective de $c$.
\vskip0.5cm
Pour que la structure (a) puisse exister, il s'av\`ere n\'ecessaire
de renforcer la notion de non-d\'eg\'en\'erescence. Pr\'ecisons tout
de suite ce fait~:

\subsection{Notion de courbe de Cartan.}

{\bf D\'efinition~:}
Une courbe $c\in {\cal C}(I\!\!R P^2)$ sera dite une {\it courbe de
Cartan} si elle ne comporte aucun point sextactique.

\vskip0.3cm

{\bf Remarque~:}
Il est possible de donner une version plus topologique de cette
d\'efinition~: une courbe de Cartan est en position g\'en\'erale, au
voisinage de chacun de ses points, par rapport \`a sa tangente, mais
aussi par rapport \`a sa {\it c\^onique osculatrice}.

Nous noterons ${\cal C}_c(I\!\!R P^2)$ (resp. $\Gamma_c(I\!\!R P^2)$
dans le cas des courbes param\'etr\'ees) l'espace de toutes les
courbes de Cartan {\it \'equivariantes}. On notera \'egalement
$\overline{\cal C}_c(I\!\!R P^2)$ (resp. $\overline\Gamma_c(I\!\!R
P^2)$) ces m\^emes espaces modulo \'equivalence projective.

Comme cons\'equence du lemme 2.6.7, nous obtenons~: l'op\'erateur de
monodromie d'une courbe de Cartan \'equivariante ne poss\`ede aucune
valeur propre \'egale \`a l'unit\'e.

\subsection{Le crochet de poisson - D\'emonstration du th\'eor\`eme
2.}

Les deux structures (a) et (b) sur une courbe de Cartan $c\in {\cal
C}_c(I\!\!R P^2)$ permettent de d\'efinir une application~:
$$\Pi\;: {\cal C}_c (I\!\!R P^2) \rightarrow {\cal P}(S^1)$$
(Nous utilisons le fait que ${\cal P}(S^1)$ est un espace affine~;
cf. 2.4.3 et 3.1.1). Cette application est $PGL(3\;;I\!\!R)$-
invariante~; elle descend sur l'espace $\overline{\cal C}_c(I\!\!R
P^2)$. On obtient l'application~:
$$\overline\Pi\;: \overline{\cal C}_c(I\!\!R P^2)\rightarrow {\cal
P}(S^1)$$

La longueur et la courbure projective sont des invariants
diff\'erentiels projectifs~: une courbe de $I\!\!R P^2$ est
uniquement d\'efini -- modulo $PGL(3\;; I\!\!R)$-- par sa courbure
$\kappa(\sigma )$ (o\`u $\sigma $ est le param\`etre projectif de la
courbe). Cette remarque montre que $\overline\Pi$ est injective.
\vskip 0.3cm
L'espace $\overline{\cal C}_c(I\!\!R P^2)$ peut alors \^etre muni
d'une structure de Poisson induite, via le plongement
$\overline\Pi$, par le crochet de Poisson sur ${\cal P}(S^1)$ (cf.
Chap. 2.7.5). Le th\'eor\`eme 2 est d\'emontr\'e.$\;\;\Box$
\vskip 0.3cm
Les feuilles symplectiques de la vari\'et\'e de Poisson
$\overline{\cal C}_c(I\!\!R P^2)$ sont constitu\'ees des courbes
$\overline c$ v\'erifiant~:
$$T(\overline\Pi(\overline c)) = C^{te}.$$

\noindent{\bf Remarque~:} La d\'efinition de ce crochet de Poisson
peut en fait se g\'en\'eraliser au cas d'une surface localement
projective quelconque.
\vfill\eject

\section{FEUILLETAGE SYMPLECTIQUE}

Il existe sur l'espace ${\cal C}(I\!\!R P^2)$ un feuilletage dont
chaque feuille est munie d'une structure symplectique. Presque toute
feuille est de codimension finie.
\vskip 0.3cm
{\bf D\'efinition.~}
Rappelons qu'une courbe projective ferm\'ee $c\in {\cal
C}_{id}(I\!\!R P^2)$ poss\`ede au moins 6 points sextactiques (cf.
2.6.6). Presque toute courbe $c\in {\cal C}(I\!\!R P^2)$ poss\`ede
un nombre fini de tels points~: $c$ est alors dite {\it
g\'en\'erique}.
\vskip 0.3cm
Soit $c\in {\cal C}(I\!\!R P^2)$ une courbe g\'en\'erique et
$\{x_1,\ldots, x_{2k}\}$ l'ensemble de ses points sextactiques
(c'est-\`a-dire $\{x_1,\ldots,x_{2k}\}$ est l'ensemble des z\'eros
de la diff\'erentielle cubique $w_3$ sur $c$). On associe alors \`a
$c$ un point

$t = t(c) \in I\!\!R P^{2k-1}$

de la mani\`ere
suivante~:
$$\left\{\begin{array}{ll}
t\;\mbox{ est la droite vectorielle de\ } I\!\!R^{2k}\;\mbox{ de
vecteur directeur }\left[\begin{array}{c}
t_1\\
\vdots\\
t_{2k}\end{array}\right]\\
\noalign{\vskip 2mm}
t_i\;:=\dsp\int^{x_{i+1}}_{x_i}w_3(x)^{\frac{1}{3}}dx\;\;\;\; \forall
i\in \{1,\ldots, 2k\}\end{array}\right.$$

Soit $T$ une classe de conjugaison dans $PGL(2\;; I\!\!R)$. Notons
${\cal M}_{T,t}$ le sous-espace de ${\cal C}(I\!\!R P^2)$
constitu\'e des courbes $c$ v\'erifiant~:
$$t(c) = t\;\;\;\mbox{ et }\;\;\; T(\kappa_c) = T$$
($\kappa_c$ est la courbure projective sur $c$, consid\'er\'ee comme
structure projective).
\vskip 0.3cm
On pose aussi $\overline{\cal M}_{T,t}\;:=\; {\cal
M}_{T,t}\biggm/{\displaystyle PGL(3\;; I\!\!R)}$.

\begin{proposition}Il existe sur $\overline{\cal M}_{T,t}$
une structure symplectique.
\end{proposition}

\noindent{\bf Preuve~:} Commen\c cons par d\'efinir une application~:
$$\overline\Pi\;:\;\overline{\cal M}_{T,t} \rightarrow{\cal
P}_T(S^1)\;\;\;\eqno{(9)}$$

Sur toute courbe $c\in {\cal C}(I\!\!R P^2)$, il existe une
structure projective~: celle d\'etermin\'ee par la courbure
projective de la courbe $c$. La longueur d'arc projective sur $c$ ne
d\'efinit pas, par contre, une structure affine sur $c$ si la
diff\'erentielle cubique $w_3$ a des z\'eros.

Soit $c$ et $c'\in {\cal M}_{T,t}$ deux courbes, $w_3$ et $w'_3$ les
diff\'erentielles cubiques associ\'ees respectivement \`a $c$ et \`a
$c'$. On peut alors identifier ces deux courbes.

Plus pr\'ecis\'ement, on peut trouver un diff\'eomorphisme $g\;:
c\rightarrow c'$ tel que~:
$$g^\ast (w'_3) = C^{te} \times w_3$$

Ce diff\'eomorphisme est \'evidemment projectivement invariant.

{\it Fixons une courbe} $c'\in {\cal M}_{T,t}$ et une {\it structure
projective} quelconque $\sigma '$ sur $c'$.

\noindent Toute courbe $c\in {\cal M}_{T,t}$ est alors munie d'une
structure projective, image r\'eciproque par le diff\'eomorphisme
$g$ de la structure $\sigma '$. Nous disposons donc de deux
structures projectives sur $c$, ce qui permet de d\'efinir une
application~:
$$\Pi\;: {\cal M}_{T,t} \rightarrow {\cal P}_T(S^1)$$
$\Pi$ \'etant invariante sous l'action de $PGL(3\;; I\!\!R)$, elle
descend sur $\overline{\cal M}_{T,t}$ et l'application
$(\diamondsuit)$ est bien d\'efinie.
\vskip0.3cm
Notons alors $\Omega \;:=\overline\Pi^\ast \omega $ o\`u $w$ est la
forme symplectique sur ${\cal P}_T(S^1)$ (cf. 2.7). Nous terminerons
la d\'emonstration par le~:

\begin{lemme}

(i) $\overline\Pi$ est bijective

(ii) $\Omega $ est ind\'ependante du choix de $c'$ et
$\sigma '$.
\end{lemme}

\noindent{\bf Preuve du lemme~:}
les deux applications $\overline\Pi$ correspondantes \`a deux choix
distincts pour $c'$ et $\sigma '$ diff\'erent par une translation
sur l'espace ${\cal P}_T(S^1)$. Or, la 2-forme $\omega $ est
invariante relativement aux translations (cf.
2.7).$\;\;\;\bullet\;\;\Box$

{\bf Remarque}

Sur l'espace $\overline{\cal C}_c(I\!\!R P^2)$ des courbures de
Cartan, la structure symplectique $\Omega $ co\"{\i}ncide avec la
structure symplectique d\'efinie sur les feuilles symplectiques (cf.
4.2).
\vfill\eject

\section{CROCHET DE GEL'FAND-DIKII}

Le crochet de Poisson de Gel'fand-Dikii (voir [G-D], [D-S]) est une
des plus int\'eressantes structures de Poisson en dimension infinie.
Les travaux s'y rattachant sont surtout populaires dans la
litt\'erature physique.

L'alg\`ebre de Poisson engendr\'ee par ce crochet joue le r\^ole de
l'alg\`ebre des sym\'etries (voir par exemple [I-Z-F]).

Du point de vue g\'eom\'etrique, l'int\'er\^et du crochet de
Gel'fand-Dikii est bas\'e sur ses propri\'et\'es naturelles
d'invariance. Il nous permettra de d\'efinir une structure
symplectique $\Diff^+(\S1)$-invariante sur l'espace des courbes
param\'etr\'eesJde $I\!\!R P^{n-1}$ modulo \'equivalence projective.

\subsection{D\'efinition du crochet de Gel'fand-Dikii}

Le crochet de Gel'fand-Dikii est d\'efini sur l'espace des
op\'erateurs diff\'erentiels lin\'eaires \`a coefficients
p\'eriodiques. Consid\'erons donc l'espace des op\'erateurs
diff\'erentiels lin\'eaires $L=\partial^n +
u_{n-2}(x)\partial^{n-2}+\ldots + u_0(x)$ dont tous les coefficients
$u_i$ sont $2\pi$-p\'eriodiques et notons le ${\cal L}_0$.

Nous allons associer \`a chaque fonctionnelle lin\'eaire $F\;: {\cal
L}_0\rightarrow I\!\!R$ un champ de vecteurs $\xi (F)$ sur ${\cal
L}_0$. Pour cette construction, nous avons besoin de quelques
r\'esultats concernant l'espace ${\cal L}_0$.

Tout d'abord, il suffira de se limiter aux fonctionnelles
lin\'eaires {\it r\'eguli\`eres} sur ${\cal L}_0$, c'est-\`a-dire
les fonctionnelles $F$ de la forme~:
$$F(L) = \sum^{n-2}_{i=0}\int_{S^1}f _i(x) u_i(x) dx$$
o\`u $f_i\in C^\infty(S^1)\;\;\;\forall i\in \{0,1,\ldots,n-2\}$
\vskip 0.3cm
{\bf D\'efinition~:}
La somme formelle~:
$$S = \sum^{n}_{i=1} a_i(x) \partial^{-i}$$
o\`u les
$a_i\in
C^\infty(S^1)\;\;,\;\;\partial^{-1}=\left(\frac{d}{dx}\right)^{-1}$
est appel\'ee un {\it symbole pseudo-diff\'erentiel} d'ordre $n$.
\vskip0.5cm A chaque symbole $S$, nous associons une
fonctionnelle lin\'eaire
$$\ell _s\;: {\cal L}_0 \rightarrow I\!\!R$$
par la construction suivante. Si $L\in {\cal L}_0$, le produit $S.L$
peut se r\'e\'ecrire comme une s\'erie formelle~:
$$S.L = \sum^{-\infty}_{i=n-1} b_i(x)\partial^i$$
Il suffit d'utiliser {\it l'identit\'e de Leibnitz}~:
$$\partial^{-1}a(x) = a(x) \partial^{-1} - a'(x)\partial^{-2} +
a''(x)\partial^{-3} -\ldots$$
(Cette identit\'e se d\'eduit de l'\'egalit\'e~:
$\partial.\partial^{-1} a(x) \equiv a(x)$.)

D\'efinissons alors~:
$$\ell _S(L)\;:= \int_{S^1} \mbox{r\'es}(S.L)(x) dx$$
o\`u $\mbox{r\'es}(S.L)\;:= b_{-1}$
Le fait suivant est alors \'evident~:

\begin{lemme} Toute fonctionnelle lin\'eaire r\'eguli\`ere
$F$ peut s'\'ecrire~:
$$F = \ell _S$$
o\`u $S$ est un symbole pseudo-diff\'erentiel.
\end{lemme}

{\bf Remarque importante~:}
Il est \'evident que la fonctionnelle $\ell _S$ ne d\'epend pas du
dernier coefficient $a_n$ du symbole $S$. Par exemple, dans le cas
$n=2$, le symbole $S=a_1\partial^{-1}+ a_2\partial^{-2}$ admet pour
fonctionnelle associ\'ee~:
$$\ell _S(\partial^2 + u) = \int_{S^1}a_1(x) u(x) dx$$

{\bf D\'efinition de l'op\'erateur hamiltonien~:}
A chaque symbole $S$, on associe un op\'erater lin\'eaire ${\cal
L}_0\rightarrow{\cal L}_0$
$$L \longmapsto L_S = L (SL)_+ - (LS)_+ L$$
(o\`u le signe + signifie que l'on consid\`ere uniquement la partie
diff\'erentielle de l'expression, c'est-\`a-dire tous les degr\'es
$\geq 0$).
\vskip0.3cm
A priori $L_S$ est un op\'erateur diff\'erentiel d'ordre $2n-1$,
mais il est facile de voir que les coefficients des op\'erateurs
$\partial^i$ sont identiquement nuls lorsque $i > n-1$.

\begin{proposition}[{[D-S]}]
A toute fonctionnelle lin\'eaire $\ell $ correspond un unique
symbole $S$ tel que~:

(i) $\ell  = \ell _S$

(ii) $\mbox{ord}(L_S) = n-2$
\end{proposition}

\noindent{\bf Preuve~:}
Le choix du coefficient $a_n$ symbole $S$ est fix\'e par la seconde
condition.$\;\;\Box$
\vskip0.3cm
{\bf Exemple $(n=2)$}
Si $\displaystyle \ell (\partial^2 + u)= \int_{S^1} f(x) u(x) dx
\;\;\;\mbox{ alors }\;\;\ell = \ell _S\;\;\mbox{ avec }\;\; S = u(x)
\partial^{-1} -\frac{u'(x)}{2} \partial^{-2}$. L'op\'erateur $L_S$ peut \^etre
interpr\'et\'e dans ce cas comme un
vecteur tangent \`a l'espace ${\cal L}_0$.
\vskip0.5cm

\begin{theoreme}[{[G-D], [D-S]}]
L'application $\xi \;: {\cal L}^\ast_0\rightarrow {\cal L}_0$
d\'efinie par~:
$$\xi (\ell) = L_S$$
(o\`u $S$ v\'erifie les 2 conditions de la proposition
pr\'ec\'edente) d\'efinit un crochet de Poisson sur l'espace ${\cal
L}_0$, c'est-\`a-dire une structure d'alg\`ebre de Lie sur l'espace
des polyn\^omes diff\'erentiels.
\end{theoreme}

{\bf D\'efinition~:} Le crochet de Poisson du th\'eor\`eme
ci-dessus est appel\'e crochet de Gel'fand-Dikii.
\vskip0.3cm
L'expression de ce crochet sur les fonctionnelles lin\'eaires est
donn\'ee par la formule suivante~:
$$\{\ell _S, \ell _T\}\;:= \ell _T(L_S)$$

{\bf Remarques~:}
(i) $n\geq 3$ Dans ce cas, le crochet de deux fonctionnelles
lin\'eaires est une fonctionnelle quadratique.
\vskip0.3cm
(ii) $n=2$ Les fonctionnelles lin\'eaires forment une alg\`ebre de
Lie isomorphe \`a l'alg\`ebre de Virasoro.
\vskip0.3cm
(iii) D'un point de vue g\'eom\'etrique, le crochet de
Gel'fand-Dikii appara\^{\i}t comme tr\`es naturel si on remarque les
deux propri\'et\'es suivantes~:
\vskip0.3cm
(1) Il est invariant relativement \`a l'action de
$\Diff^+(\S1)$.
\vskip0.3cm
(2) La restriction de ce crochet aux fonctionnelles
$F(u_{n-2})$ d\'efinit un crochet d'alg\`ebre de Lie. Cette
alg\`ebre de Lie co\"{\i}ncide avec l'alg\`ebre des fonctions sur
les op\'erateurs de Hill~:
$$L = \partial^2_x + \frac{6}{n(n-1)(n+1)} u_{n-2}(x)$$
(voir [Kh]).

Consid\'erons, de plus, l'action infinit\'esimale de l'alg\`ebre de
Lie $Vect(S^1)$ associ\'ee \`a l'action de $\Diff^+(\S1)$. Cette action
infinit\'esimale conserve le crochet de Gel'fand-Dikii.
L'application moment correspondant \`a cette action est donn\'ee par~:
$$\hbox{(courbe dans} I\!\!R P^{n-1})\longmapsto \hbox{ (courbure
projective associ\'ee)}$$

\subsection{Structure symplectique sur l'espace des courbes
projectives param\'etr\'ees}

On \'etudie ici les feuilles symplectiques du crochet de
Gel'fand-Dikii.

Le r\'esultat principal dans [K-O] est le~:

\begin{theoreme} Le crochet de Gel'fand-Dikii d\'efinit une structure
symplectique
sur chaque sous-espace ${\cal L}_T \subset {\cal L}_0$ constitu\'e
par les op\'erateurs \`a monodromie\footnote{Rappelons que $T$ d\'esigne
en fait une classe de conjugaison dans le groupe projectif
$PGL(n\;;I\!\!R)$} fix\'ee.
\end{theoreme}

En d'autre termes, chaque espace $\overline\Gamma_T(I\!\!R
P^{n-1})\cong {\cal L}_T$ (cf . 2.4) est une {\it vari\'et\'e
symplectique}.
\vskip0.3cm
\noindent{\bf Preuve.}
Il nous faut d\'emontrer deux faits~:

(a) Un champ hamiltonien $L_X$ est tangent \`a ${\cal L}_T$.
\vskip0.5cm
(b) Tout vecteur tangent \`a ${\cal L}_T$ est donn\'e par un champ
hamiltonien.

La donn\'ee d'un op\'erateur $L$ est \'equivalente \`a celle de
l'espace des solutions de l'\'equation diff\'erentielle associ\'ee
$Ly =0$. A un symbole $S$ correspond un champ de vecteurs
$L\longmapsto L_S$. Ce symbole induit \'egalement un champ sur
chaque espace de solutions $E_L =\{y\in C^{\infty }(I\!\!R) \vert Ly = 0\}$~:
$$y \longmapsto y_S$$
Le r\'esultat est~:
$$y_S = - (SL)_+ y\eqno {(10)}$$
En fait, par d\'efinition, on a l'identit\'e suivante~:
$$L_S( y) + L(y_S) =0$$
(a) est alors \'evident en consid\'erant (10) car $y\mapsto y_S$ est
lin\'eaire en $y$.

R\'eciproquement, si on fixe $y_S$, le symbole $S$ s'obtient comme
solution d'un syst\`eme d'\'equations lin\'eaires. On v\'erifie
facilement que les coefficients $a_i(x)$ de $S$ sont
$2\pi$-p\'eriodiques si et seulement si la monodromie est
fix\'ee.$\;\;\;\Box$
\vskip0.3cm
{\bf Exemple~: $(n=3)$~}
L'espace $\overline\Gamma_{Id}(I\!\!R P^2)$ est constitu\'e de 3
feuilles symplectiques. Chacune de ces feuilles correspond \`a une
des 3 classes d'homotopie dessin\'ees \`a la Fig. 1 (Chap.
2.3.2).

\subsection{Formules explicites pour les formes symplectiques}

Nous pr\'esentons dans ce chapitre la formule donnant la forme
symplectique induite par le crochet de Gel'fand-Dikii dans un cas
particulier.
\vskip0.3cm
{\bf D\'efinition~:}
Une courbe $c\in {\cal C}(I\!\!R P^{n-1})$ sera dite {\it non
oscillante} si tout hyperplan projectif de $I\!\!R P^{n-1}$
rencontre la courbe en au plus $n-1$ points distincts.
\vskip0.3cm
Si $\gamma(x)$ est une courbe param\'etr\'ee non oscillante $\in
\Gamma(I\!\!R P^{n-1})$, alors l'op\'erateur diff\'erentiel
associ\'e $L_\gamma$ (cf. 2.4) est non oscillant au sens classique
. C'est-\`a-dire, chaque solution $y$ de l'\'equation
diff\'erentielle associ\'ee $L_\gamma. y = 0$ admet au plus $n-1$
z\'eros sur l'axe r\'eel.
\vskip0.3cm
Le cas particulier pour lequel nous allons expliciter la forme
symplectique de Gel'fand-Dikii est donc celui des courbes
\'equivariantes, non oscillantes et \`a monodromie fix\'ee.
\vskip0.3cm
Pr\'esentons tout d'abord deux r\'esultats classiques de la
th\'eorie des \'equations diff\'eentielles lin\'eaires, r\'esultats
qui nous seront utiles pour la suite~:

\begin{lemme} (Sturm)~:

Un op\'erateur $L\in {\cal L}^n$ est non-oscillant si et seulement
si on peut trouver $n-1$ solutions $y_1,\ldots, y_{n-1}$ pour
l'\'equation $Ly =0$ telles que les $n-1$ fonctions~:
$$y_1\blanc,\blanc w_1 =\left|\begin{array}{cc}
y_1 & y_2\\
y'_1 & y'_2\end{array}\right|,\blanc \ldots\blanc,\blanc w_{n-1}
=\left|\begin{array}{ccc}
y_1 &\ldots &y_{n-1}\\
\noalign{\vskip 2mm}
\vdots &&\\
\noalign{\vskip 2mm}
y^{n-2}_1&\ldots& y^{n-2}_{n-1}\end{array}\right|$$
ne s'annulent en aucun point.
\end{lemme}
\begin{lemme} (Frobenius)~:

Un op\'erateur $L\in {\cal L}^n$ est non-oscillant si et seulement
si $L$ poss\`ede une factorisation en $n$ op\'erateurs d'ordre 1~:
$$L = (\partial + a_1 +\ldots + a_{n-1})(\partial - a_{n-1})\ldots
(\partial - a_1)$$
o\`u les $a_i\in C^\infty(I\!\!R)\;\;\;\forall i\in \{1,\ldots,
n-1\}$.
\end{lemme}
Les coefficients $a_i$ ci-dessus peuvent \^etre calcul\'es en
posant~:
$$\left\{\begin{array}{ll}
\displaystyle
a_1 = -\frac{y'_1}{y_1}\\
\noalign{\vskip 2mm}
\displaystyle
a_i = - a_{i-1}-\frac{w'_i}{w_i}\;\;\;\;\forall i\in \{2,\ldots,
n-1\}\end{array}\right.$$
o\`u $y_1$ et $w_i$ sont les fonctions vues dans le lemme 7.1.
\vskip0.3cm
{\bf Remarque~:}
Supposons que les coefficients $u_i$ dans l'op\'erateur $L$ soient
$2\pi$-p\'eriodiques. On peut alors trouver une factorisation de $L$
o\`u les coefficients $a_i$ sont \'egalement $2\pi$-p\'eriodiques. En
fait, il suffit que l'op\'erateur de monodromie $T$, associ\'e \`a $L$,
soit repr\'esent\'e par une matrice triangulaire dans la base
$\{y_1,\ldots, y_{n-1}, y\}$.
\vskip0.3cm
Soit $\gamma$ une courbe param\'etr\'ee $I\!\!R\rightarrow I\!\!R
P^{n-1}$ non oscillante et \'equivariante. Notons $\delta \gamma$ et
$\widehat\delta  \gamma$ deux d\'eformations infinit\'esimales de
$\gamma$ (en d'autres termes, deux vecteurs tangents en $\gamma$ \`a
l'espace des courbes non oscillantes).
On peut
identifier $\delta \gamma$ (resp. $\widehat\delta \gamma$) \`a $\delta  a
= (\delta a_1, \ldots,\delta a_{n-1})$ (resp. \`a $\widehat\delta a
=(\widehat\delta a_1,\ldots,\widehat\delta a_{n-1})$).

\vskip0.3cm
Soit $A = (A_{ij})$ la matrice carr\'ee d'ordre $n-1$ d\'efinie par~:
$$A_{ij}\;:=\;\delta _{ij} - \frac{1}{n}$$
On a alors le~:

\begin{theoreme}

La forme symplectique $\omega $ de Gel'fand-Dikii sur l'espace des
courbes non oscillantes \`a monodromie fix\'ee est donn\'ee par~:
$$\omega (\delta \gamma, \widehat\delta \gamma) =
\int_{S^1}( \delta a A \partial^{-1} \overline{\delta a})
(x) dx$$
\end{theoreme}
\vfill\eject

{\parskip12pt{\bf R\'ef\'erences}

[Bo] - G. Bol - ``Projektive Differentialgeometrie'' -
Studia Mathematica / Mathematische Lehrb\"ucher - Vandenhoeck $\&$
Ruprecht - G\"ottingen - 1950.

[Br]  - J.L. Brylinski - ``Loop spaces, Charateristic Classes
and Geometric Quautization'' - Progress in Math. - Vol {\bf 107} -
Binkh\"auser - 1992.

[Bu]  - S. Buchin - ``Affine Differential Geometrie'' - Gordon
$\&$ Breach, Sci. Publ., inc., New-York - Science Press - Beijurg
- 1983.

[Ca]  - E. Cartan - ``Le\c cons sur la th\'eorie des espaces
\`a connexion projective'' - Gauthier -Villars - Paris - 1937.

[C-F] - E. C$\check{e}$ch $\&$ G. Fubini - ``Introduction
\`a la g\'eom\'etrie projective diff\'erentielle des surfaces'' -
Gauthier-Villars - Paris - 1931.

[Ch] - S. Chihi - ``Sur les feuilletages de codimension 1
transversalement homographiques'' - Th\`ese - publ. de l'IRMA -
Strasbourg - 1979.

[D-S] - V.G. Drinfel'd $\&$ V.V. Sokolov - ``Lie algebras
and equations of Korteweg - De Vries type'' - J. Soviet Math. - Vol. {\bf
30} - 1985 - p. 1975 \`a 2036.

[F-I-Z]  - Ph. Di Francesco, C. Itzykson $\&$ J.B. Zuber -
``Classical W-algebras'' - Saclay SPhT 90 / 149 - Princeton PUPT n.1211 -
1990.

[G-D]  - I.M. Gel'fand $\&$ L.A. Dikii - ``A family of
hamiltonian structures connected with integrable nonlinear differential
equations'' - in~: I.M. Gel'fand collected papers (S.G. Gindikin et al,
eds) - Vol. {\bf 1} - Springer-Verlag - 1987 - p. 625 \`a 646.

[G-F]  - I.M. Gel'fand $\&$ D.B. Fuchs - ``The cohomologies
of the Lie algebra of the vector fields in a circle'' - Func.
Anal. Appl. - Vol. {\bf 2} - n.4 - 1968 - p. 342 \`a
343.

[Go] - W. Goldman - ``Geometric structures on manifolds and
varieties of representations'' - Contemporary Math. - Vol. {\bf 74}
- 1988 - p. 169 \`a 198.

[Gr]  - M. Green - ``The moving frame, differential invariants
and rigidity theorems for curves in homogeneous spaces'' - Vol. {\bf 45}
- n.4 - Duke Math. J. - 1978 - p. 735 \`a 779.

[Ha]  - G. Halphen - ``Sur les invariants differentiels'' -
Th\`ese, Paris, 1878 - in~: Oeuvres - II - p. 197 \`a 257.

[Hu]  - J. Hubbard - ``The monodromy of projective structures''
 - in : Riemann surfaces and related topics ; Stony Brook 1978 - Eds : I. Kra
\& B. Maskit
- Annals of Math. Stud. - Vol. {\bf 97} - 1981 - p. 257 \`a 275.

[Ig]  - P. Iglesias - ``La trilogie du moment'' - Indag. Math.
(\`a para\^{\i}tre)

[Iv]  - N.V. Ivanov - ``Projective structures, flat burdles
and K\"ahler metrics on moduli spaces'' - Math. USSR Sbornik - vol {\bf
61} - n.1 - 1988 - p. 211 \`a 224.

[K-O]  - B. Khesin $\&$ V. Yu. Ovsienko - ``Symplectic leaves
of the Gel'fand-Dikii brackets and homotopy classes of nondegenerate
curves'' - Func. Anal. Appl. - Vol. {\bf 24} - n.
1 - 1990.

[K-S] - B. Khesin $\&$ B. Shapiro - ``Non degenerate curves
on $S^2$ and orbit classification of the Zamolodchikov algebra'' -
Comm. Math. Phys. - Vol. {\bf 145} - 1992 - p. 357
\`a 362.

[Kh]  - T.G. Khovanova - Lie algebras of Gel'fand-Dikii and
the Virasoro algebra, Funk. Anal. iprilozhen., Vol. {\bf 20}, n.4, - 1987
- p. 89 \`a 90.

[Ki 1]  - A.A. Kirillov - ``Infinite dimensional Lie
groups~: their orbits, invariants and representations. The geometry of
moments'' - Lect. Notes in Math. - Vol.{\bf 970} -
Springer-Verlag - 1982 - p. 101 \`a 123.

[Ki 2]  - A.A. Kirillov - ``Orbits of the group of
diffeomorphisms of a circle and local Lie superalgebras'' - Func.
Anal. Appl. - Vol. {\bf 15} - n. 2 - 1981 - p. 135 \`a
137.

[Ki-Y]  - A.A. Kirillov $\&$ D.V. Yuriev - ``Representations
of the Virasoro algebra by the orbit method'' - Journal of Geometry and
Physics - Vol. {\bf 5} - n.3, - 1988 - p. 351 \`a 363.

[Ku] - N.H. Kuiper - ``Locally projective spaces of dimension
one'' - Michigan Mathematical Journal - Vol. {\bf 2} - 1954 - p. 95 \`a
97.

[L-P]  - V.F. Lazutkin $\&$ T.F. Pankeratova - ``Normaf forms
and versal deformations for Hill's equations'' - Func. Anal.
Appl. - Vol. {\bf 9} - n.4 - 1975 - p. 306 \`a 311?

[Li 1] - J.A. Little - ``Nondegenerate homotopies of curves
on the unit 2-sphere'' - J. Diff. Geom. - Vol. {\bf 4}
- n. 3 - 1970 - p. 339 \`a 348.

[Li 2]  - J.A. Little - ``Third order nondegenerate
homotopies of the space curves'' - J. Diff. Geom. - Vol. {\bf 5} - 1971 -
p. 503 \`a 515.

[O] - V.Yu. Ovsienko - ``Classification of third-order linear
differential equations and symplectic sheets of the Gel'fand-Dikii
bracket'' - Math. Notes - Vol.{\bf 47} - n.5-6 - Nov. 1990 - p. 465 \`a
470.

[Se 1] - G.B. Segal - ``Unitary representations of some
infinite dimensional groups'' - Comm. Math. Phys. - Vol. {\bf 80} - n.3 -
1981 - p. 301 \`a 342.

[Se 2]  - G.B. Segal - ``The geometry of the KdV equation'' -
in~: Trieste Conference on topological methods in quantum field theories
- W. Nahm $\&$ al, eds - World scientific - 1990 - p. 96 \`a 106.

[Sh] - M.Z. Shapiro - ``Topology of the space of
nondegenerate closed curves'' - Mat. Sbornik (\`a para\^{\i}tre)

[S-T]  - D. Sullivan $\&$ W. Thurston - ``Manifolds with
canonical coordinate charts~: some examples'' - L'Enseign.
Math. - t. {\bf 29} - 1983 - p. 15 \`a 25.

[Wh]  - H. Whitney - ``On regular closed curves in the plane''
Compositio Math. - {\bf 4} - 1937 - p. 276 \`a 284.

[Wi]  - E.J. Wilczynski - ``Projective differential geometry
of curves and ruled surfaces'' - Leipzig - Teubner - 1906.

[Ws]  - G. Wilson - ``on the Adler-Gel'fand-Dikii brucket'' -
in~: Proc. of the CRM workshop on hamiltonian systems, - J. Harnad $\&$
J.E. Marsden, eds - Publ. CRM - 1990 - p. 77 \`a 85.

[Wu]  - T. Wurzbacher - ``Symplectic geometry of the loop
space of a riemannian manifold'' - SFB 237 Preprint - Ruhr. Universit\"at
Bochum - 1993.
}

\end{document}